\documentclass[a4paper,12pt,oneside,onecolumn]{article}
\usepackage{color,graphicx}

\usepackage{geometry}%
\usepackage{amsmath,amssymb}             
\usepackage{cite}
\usepackage{bm}

\definecolor{c1}{rgb}{1,0,0}
\definecolor{c2}{rgb}{0,0,1}
\def\COM{.}
\newcommand{\ppd}[2]{\frac {\partial #1 }{\partial #2}}

\newcommand{\DD}[2]{\frac {D #1 }{D #2}}
\newcommand{\pdd}[2]{\frac {\partial^2 #1 }{\partial {#2}^2}}

    \usepackage[tight]{subfigure}
 \newcommand{\etal}{\textit{et al. }}
 \newcommand{\ve}[1]{\left\{ {\bf #1} \right\}}
   \newcommand{\ten}[1]{\left[ {\bf #1} \right]}

\newcommand{\ie}{, \textit{i.e.}, }
\newcommand{\eg}{, \textit{e.g.}, }

\newcommand{\figc}[1]{Fig.~\ref{#1}}

\newcommand{\secc}[1]{Sec.{ }\ref{#1}}

\newcommand{\eq}[1]{Eq.~(\ref{#1}) }

\newcommand{\eqc}[1]{Eq.~(\ref{#1})}
\newcommand{\apc}[1]{Appendix~\ref{#1}}
 \newcommand{\fo}[1]{$\mathrm{#1}$}

\usepackage[colorlinks=true,linkcolor=black,filecolor=black, citecolor=black,urlcolor=blue]{hyperref}

\title{A mesoscale granular model for the mechanical behavior of alloys during solidification}

\author{St\'ephane Vern\`ede$^{1,2}$, Jonathan A. Dantzig$^{1,3}$,
        Michel Rappaz$^1$}%
\date{}
\begin{document}

\maketitle
\begin{center}
Published in Acta Materialia, Volume 57, Issue 5, March 2009, Pages 1554-1569  \\

\href{http://dx.doi.org/10.1016/j.actamat.2008.12.006}{view at publisher}

\begin{small}
$^1$ Computational Materials Laboratory\\
Ecole Polytechnique F\'ed\'erale de Lausanne\\
Station 12, Lausanne, CH-1015 Switzerland\vspace{5 mm }\\
$^2$Alcan Centre de Recherches de Voreppe,\\ ZI Centr'Alp, 725 rue Aristide Berg\`es\\
 BP 27, Voreppe, FR-38341 France \vspace{5mm}\\
$^3$ Department of Mechanical Science and Engineering \\
University of Illinois, 1206 West green Street\\
Urbana, IL 61801 USA
\end{small}
\end{center}
\begin{center}
keywords: Solidification, Percolation, Micromechanical Modeling, Mesostructure, Hot Cracking
\end{center}

\begin{abstract}

We present a two-dimensional granular model for the mechanical behavior of
an ensemble of globular grains, during solidification. The grain structure
is produced by a Voronoi tessellation based on an array of predefined
nuclei. We consider the fluid flow caused by grain movement and
solidification shrinkage in the network of channels that is formed by the
faces of the grains in the tessellation. We develop the governing
equations for the flow rate and pressure drop across each channel when the
grains are allowed to move, and we then assemble the equations into a
global expression that conserves mass and force in the system. We show
that the formulation is consistent with dissipative formulations of
non-equilibrium thermodynamics. Several example problems are presented to
illustrate the effect of  tensile strains and the availability of liquid
to feed the deforming microstructure. For solid fractions below
$g_s=0.97$, we find that the fluid is able to feed the deformation at low
strain, even if external feeding is not permitted. For solid fractions
above $g_s=0.97$, clusters of grains with ``dry'' boundaries form, and
fluid flow becomes highly localized.

\end{abstract}

\section{Introduction}

The last stage solidification of alloys is a critical step in casting and
welding processes during which several defects can form
\cite{Campbell1991}. In addition to porosity, the most severe of these
defects is probably hot cracking\ie a spontaneous failure of the material
while it is still semi-solid. This defect, which typically occurs in
dilute and hard alloys, limits the productivity of cast houses and
restricts the range of alloys that can be produced. \cite{Commet2006} It
also severely limits the weldability of this class of alloys.

In dilute alloys, solid grains are separated by thin continuous liquid
films up to high volume fraction of solid $g_s$ (typically up to $g_s
\approx 0.95$), especially at high angle grain boundaries, where
coalescence of solid grains is made difficult by the large grain boundary
energy\cite{coal}. As stresses and strains are generated in the not yet
fully coherent solid, the presence of these liquid films together with the
low permeability of the mushy zone near $g_s=0.95$, which prevents
efficient feeding, makes the material extremely brittle. 
\cite{borland,Clyne} Deformation induced by thermal and solidification
shrinkage tends to localize at these liquid films, which act as a brittle
phase. They pose little resistance to tensile strains and, as they open,
the newly created volume cannot be fed by intergranular liquid. The so
called ``brittle range,'' meaning the $g_s$-range where the mushy zone
exhibits  low strength and low ductility, can be measured, for example, by
tensile tests on partially solidified alloys.
\cite{rev_crique,ludwig_model,Vernede2007}

Accurate prediction of hot cracking requires the knowledge of:
\begin{enumerate}
\item
The thermal history of the cast or welded part. This is usually
done using volume-averaged multi-phase heat and mass transfer models
\cite{Ni1991,simumat};
\item
Realistic mechanical constitutive equations for the
semi-solid alloy \cite{ludwig_model}; and
\item
Conditions under which a hot tear will form,
also known as a hot cracking criterion \cite{rdg,Grandfield2005,Magnin}.
\end{enumerate}
Hot tearing models based on such data have been implemented successfully
in casting simulation codes by several authors.
\cite{Mathier2006,Monroe2005,MHamdi2006}

These existing models treat the material as a continuum whose properties
are represented by volume-averaged quantities at a length scale that is
large compared to the microstructure, and small  compared to the
variations of macroscopic fields. By their very construction, they cannot
include the important transitions in microstructure at high volume
fraction solid, such as the localization of liquid films to the periphery
of increasingly large grain clusters, as described by percolation theory.
\cite{percotheorie} Thus, the loss of locality of the length scale implies
that no appropriate representative volume element can be defined. This
limits the utility of such models, because it is in precisely this
transition region that hot cracks form, as recently demonstrated
experimentally by Gourlay and Dahle, emphasizing the importance of the
granular nature of solidifying alloys. \cite{Gourlay2007,Martin2007}

In order to go beyond the limits imposed by the volume-averaged
methods, in this work we adapt \textit{granular} or \textit{discrete
element models} (DEM), which have been developed in the context  of
the mechanics of granular materials.
\cite{Ferrez2001,Pournin2005,Martin2006} These models  simulate the
behavior of a large number of spherical grains, which may be either
rigid \cite{Ferrez2001,Pournin2005} or deformable \cite{Martin2006}.
They consider the interactions between the grains due to solid-solid
contact but neglect the influence of the surrounding medium. To
extend these models to the case of semi-solid alloys at high solid
fraction, we must consider not only the solidification of each
grain, but also the influence of fluid flow on the mechanical
behavior of the mushy zone.

The first model using the granular approach for the solidification
and coalescence of globular grains in 2D was proposed by Mathier
\etal \cite{vince2} This work was further developed by two of the
present authors. \cite{troisieme} In \cite{second}, a percolation
analysis was presented that identified the various transitions in
the mushy zone that appear naturally in this approach, as well as a
model for liquid feeding. However, all these contributions made the
major assumption that the grains remained fixed. In the present
contribution, we derive a 2D granular mechanical model for globular
microstructure that includes both the flow of intergranular liquid
and the displacement of solid grains.

\section{Solidification model}

The 2D mechanical model that we present here is based on the
solidification model presented in detail in previous publications.
\cite{vince,troisieme} The model is appropriate for inoculated alloys
whose final grain structure is fine and globular. For the sake of
completeness, we briefly recall the main features of this solidification
model before presenting the mechanical DEM approach.

\begin{figure}
\begin{center}
\includegraphics[width=0.6\textwidth]{\COM/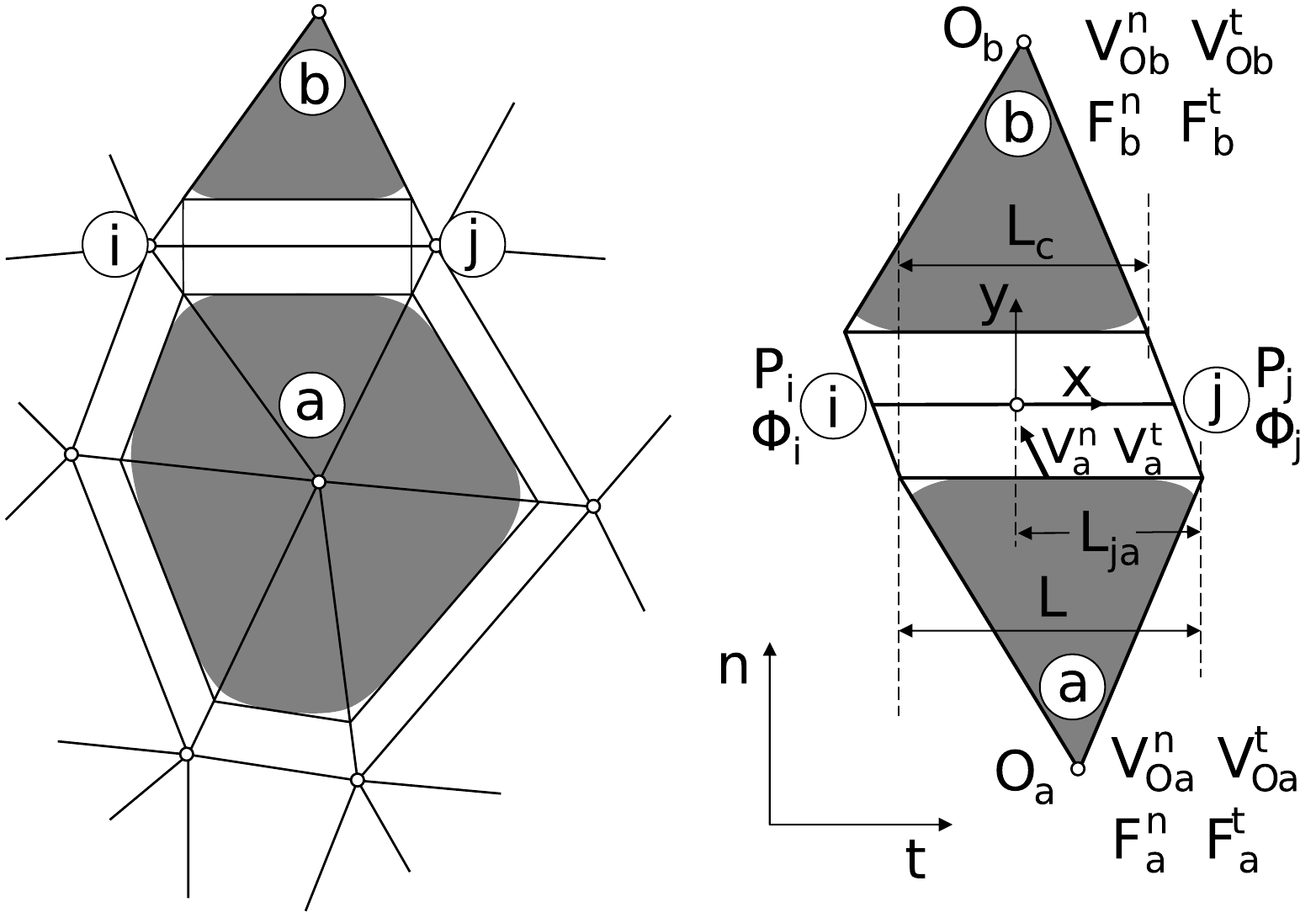}
\caption{Left, grain as computed with the Voronoi tessellation
method. Right, notations for a liquid channel used as the basic
element of the mechanical model.} \label{solidif}
\end{center}
\end{figure}

Consider a population of grains distributed randomly in space with a
specified average density. Nucleation is assumed to be instantaneous, and
the temperature difference across a typical grain is taken to be small
compared to the undercooling\ie small thermal gradient. The final grain
structure is therefore close to the Voronoi tessellation of the original
set of nuclei. \cite{charbon} Beginning with a random distribution of
nucleation centers, the Voronoi tessellation of the set is computed from
the mediatrix of the segments connecting two neighbor nucleation centers
(see Fig. \ref{solidif}). A 2D grain is then identified as an ensemble of
triangles having the nucleation center as one corner and two vertices on
the grain boundary. Computation in the model proceeds in two steps. In the
first step,  the interface of each grain during growth is approximated by
a linear segment parallel to the grain boundary. In this step, we neglect
solute flux between elementary triangles, which reduces the solute
diffusion calculation to a one-dimensional problem in each triangle. We
assume complete mixing in the liquid. Back-diffusion in the solid can be
easily incorporated in the model. \cite{vince,troisieme}. In the second
step, the sharp corners of the polyhedral grains are smoothed using a
procedure that accounts for the local curvature undercooling. 
Further details of this model, together with a
discussion of its limitations and domain of validity are given in Ref.
\cite{troisieme}.

\section{Mechanical model}
\subsection{Basic hypotheses}

We begin the mechanical model with the microstructure computed by the
solidification model. The idea is to derive a DEM for the mechanical
response of the mushy zone that retains the main physical aspects at the
scale of the grains, yet is simple enough to be computationally tractable.
To that end, the following assumptions are made:
\begin{enumerate}
\item
\emph{The grains are undeformable.} This assumption is adequate to obtain
insight into the physics of hot tearing, where deformations are small. A
more complete model of the rheology of the mushy zone should also include
deformation of the solid, which is known to be important for volume
fractions of solid $g_s \gtrsim 0.6$, known as the traction coherency solid
fraction. \cite{ludwig_model,shear_couette,Braccini2002}.
\item
\emph{Grain movement can occur only by translation.}  This
hypothesis seems very restrictive, but experimental studies have
shown that at high temperature and high solid fraction, the most
important deformation mechanism is grain boundary sliding.
\cite{rev_crique} Making this assumption greatly improves the
computational efficiency of the model, because detection of contacts
between polygonal grains is difficult if rotations are permitted.
\cite{Ferrez2001}
\item
\emph{Liquid channels smaller than a pre-defined coalescence interaction
distance $\delta$, on the order of a few nm, are taken to be fully solid.}
\cite{vince2,coal}. Using such a cut-off improves the computational
behavior of the model by eliminating very large coefficients that would
otherwise appear in the matrix of the linear system derived in
Sect.~\ref{sect:global}.
\item
\emph{The intergranular fluid is Newtonian and incompressible, and no-slip
conditions apply at the solid-liquid ($s-\ell$) interface.} We adopt this
constitutive model for the fluid even though the channels between grain
are very thin. We note that measurements by Israelachvili showed that the
viscosity observed in bulk samples is still valid for  films as thin as 5
nm. \cite{Israelachvili1986,Tabeling2003}. Another effect observed in thin
films is slipping at $s-\ell$ interfaces which may be due to the formation
of a nanometer scale air gap. \cite{Genne2002,Tabeling2003} This
phenomenon occurs when the shear stress exceeds a critical value, and
might be an interesting phenomenon to consider in hot tearing, especially
if gas porosity appears at grain boundaries. It is not included in this
work.
\end{enumerate}

\subsection{Notation}

Our model considers the liquid network formed between the grains.
The basic element is a liquid channel surrounded by two solid
grains, designated $a$ and $b$ in \figc{solidif}. The element has
four integration points. The first two points are the grain centers,
denoted $O_a$ and $O_b$ for grains $a$ and $b$, respectively. At
these points, we consider two conjugate vector quantities, the velocity of
the grain $\ve{V}_{O_a}$ (or $\ve{V}_{O_b}$) and the force exerted
by the grain on a liquid channel $\ve{F}_a$ (or $\ve{F}_b$). The two
other integration points are the ends of the liquid channel, denoted
$i$ and $j$. The conjugate quantities considered at these points are
the fluid flux $\Phi_i$ (or $\Phi_j$) and the pressure $P_i$ (or
$P_j$). Uppercase letters are used for these entities at the
integration points, whereas lowercase letters will be used for the
associated fields (e.g, $\ve{v}$ for the velocity and $p$ for the
pressure at any point in the liquid channel).

Relative translation between neighboring grains can produce a mismatch at
the extremities of the channel. With reference to \figc{solidif}, we
introduce the following length measures: $L$ is the length of the $s-\ell$
interface for one channel, $L_c$ is the length of the channel where the
two grains effectively face each other, and $L_{ja}$ is the length of the
$s-\ell$ interface from the center of the channel to the extremity of
grain $a$ near vertex $j$. We also define $L_{ia}=L-L_{ja}$. The
half-width of the channel is designated as $h$. Note also that even though
the channel can be curved at its extremities, it is modeled with straight
lines for the calculation of the pressure field.\footnote{Note that
the flux balances that we introduce in the next section with the
polyhedral envelope of the grains are strictly the same as those done for the
rounded grains. Indeed, as the fluid will be considered as incompressible,
the mass of fluid in between the grains and their polyhedral envelope
remains constant.} The special case where
relative motion of the grains causes their faces to no longer align
anywhere\ie $L_c=0$, is treated in \secc{contact}.

The volume change associated with solidification shrinkage\ie due to
the density difference ($\rho_s-\rho_l$) between the solid and
liquid phases, produces a compensating flow in the liquid near the
interface. It is most convenient to develop the expressions for the
flow in terms of the normal and tangential components of the
velocity, denoted with superscripts ``$n$'' and ``$t$'',
respectively
 \begin{equation}
\ve{V}_a =\left(
\begin{array}{c}
V^t_a \\ V^n_a\\
\end{array}
\right)
=\left(
\begin{array}{c}
V^t_{O_a} \\ V^n_{O_a}-\beta v^* \\
\end{array}
\right)
\end{equation}
where $\beta$ is the solidification shrinkage $(\beta=\rho_s/\rho_l-1)$,
and $v^*$ is the speed of the $s-\ell$ interface as given by the
solidification model. We choose the element normal vector $\ve{n}$ to
point toward the liquid from grain $a$, and the tangential vector $\ve{t}$
is directed from vertex $i$ to vertex $j$. Note that, due to this
orientation convention, the fluid velocity at the interface with grain $b$
is written:
 \begin{equation}
\ve{V}_b=\left(
\begin{array}{c}
V^t_{O_b} \\ V^n_{O_b}+\beta v^* \\
\end{array}
\right)
\label{Vb}
\end{equation}
Symmetry of the solidification model implies that the
velocity $v^*$ of the $s-\ell$ interface of grain $b$ is equal to
that of its neighbor grain $a$. The coordinates in the $\ve{t}$ and
$\ve{n}$ directions are noted $X$ and $Y$, respectively. The origin is
defined at the center of the channel (see \figc{solidif}).

\subsection{Integration of the constitutive equations}

The scaling analysis given in \apc{dim_analys} shows that the 
$X-$direction momentum balance in the channel reduces to the following
simpler relation between the pressure in the liquid channel $p(X)$
and the fluid velocity
\begin{equation}
\ppd{p}{X} =\mu\pdd{v_X}{Y}
\label{base}
\end{equation}
where $\mu$ is the dynamic viscosity. The volumetric flow rate in the
channel $\phi_{i\rightarrow
j}(X)$ from vertex $i$ to $j$ is defined by
\begin{equation}
\phi_{i\rightarrow j}(X)= \int _{-h}^{h} v_X(X,Y) dY
\label{flux_def}
\end{equation}
We also have the continuity equation for a constant density fluid in 2-D,
\begin{equation}\label{continuity}
\ppd{v_X}{X}+\ppd{v_Y}{Y}=0
\end{equation}
Integrating Eq.~\ref{continuity} over the width of the channel and using
the definition of $\phi_{i\rightarrow j}$ in Eq.~(\ref{flux_def})
gives
\begin{equation}
 \ppd{\phi_{i\rightarrow j}}{X} = -V^n
\label{pompe}
 \end{equation}
where
\begin{equation}
V^n=V^n_b-V^n_a=V^n_{O_b}-V^n_{O_a}+2\beta v^* \label{pompe.1}
\end{equation}
represents a source or sink in the channel. Note that if the velocity of
the grains is zero (or uniform)\ie $V^n=2\beta v^*$, Eqs. (\ref{base}) and
(\ref{pompe}) reduce to those given in \cite{troisieme}, where we considered
only the shrinkage-induced intergranular flow.

\begin{figure}[htb]
\begin{center}
\includegraphics[width=0.9\textwidth]{\COM/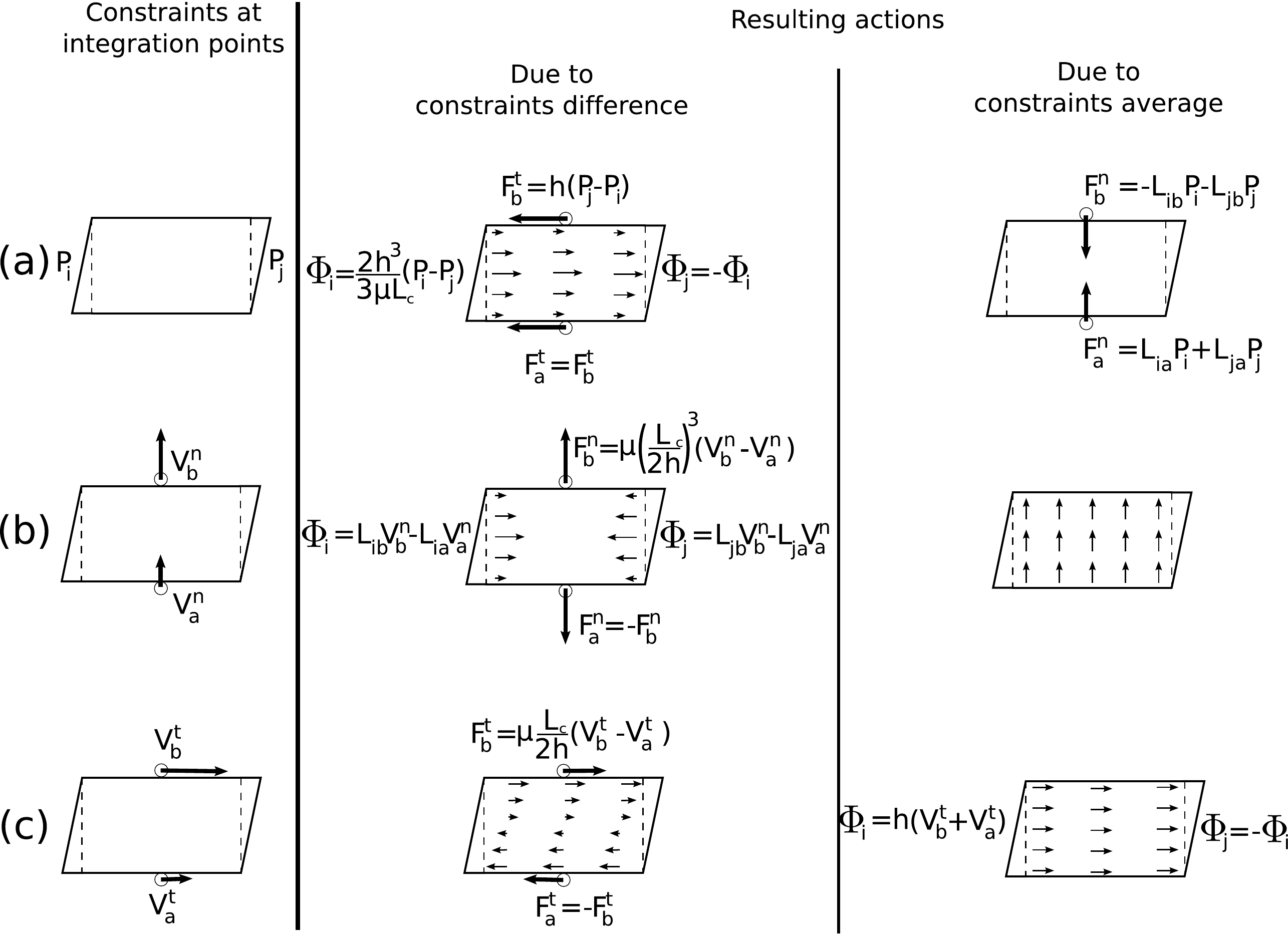}
\caption{Representation of the various individual constraints imposed
on a liquid channel and on the grains: (a) Simple flow in between two
immobile grains; (b) Normal displacement of the grains; (c) Tangential 
displacement of the grains.}
\label{elem}
\end{center}
\end{figure}

Equations (\ref{base}) and (\ref{pompe}) are integrated in
\apc{app_stress} for the general case. To provide a  better
understanding of the underlying physics, let us consider a few
special cases where the grain and/or liquid movement isolates the
individual contributions of certain important phenomena. Since the
governing equations are linear, solutions for cases that combine
these phenomena can be obtained by superposition of these simple
cases. The cases are illustrated in \figc{elem}.

Case (a), corresponding to the top line of \figc{elem}, neglects
shrinkage-induced flow ($\rho_s=\rho_l$), and considers an imposed
pressure differential between vertices $i$ and $j$. The grain
velocities are set to zero for this case. In the part of the channel
where the two $s-\ell$ interfaces face each other, the fluid
velocity profile $v_X(Y)$ has the usual parabolic profile given by
\begin{equation}
v_X(X,Y) = \frac{1}{2\mu}\ppd{p}{X} (Y^2-h^2) =
           \frac{P_i-P_j}{2\mu L_c}\ppd{p}{X} (Y^2-h^2)
\end{equation}
Thus, the volumetric flow rate at the vertices $i$ and $j$ is given by:
\begin{equation}\label{vol_flow_rate}
\Phi_i (P_i,P_j) =  \frac{2 {h}^3}{3 \mu L_c} (P_i-P_j) = - \Phi_j (P_i,P_j)
\end{equation}
This equation describes the flow between two parallel planes.  In 2D, any
contact between neighboring grains blocks all flow in the channel, whereas
in 3D the liquid can flow around the contact. To include this important 3D
feature in our 2D model, we assume that when two grains make contact, a
small ``pipe'' remains open. The radius of this pipe, $r_p$, is estimated
to be of the same order of magnitude as the radius of curvature at the
grain corner derived in \cite{troisieme} (see \figc{solidif}). Moreover,
if the solidification microstructure were to be extended in the third
dimension, there would be one such pipe each average grain diameter
$D_{av}$. Summing the flow rates for the channel and the pipe gives
\begin{equation}
\Phi_i (P_i,P_j) =  \left( \frac{2 {h}^3}{3 \mu L_c} +\frac{\pi
{r_p}^4}{8 \mu D_{av} L_c}  \right) (P_i-P_j) = - \Phi_j (P_i,P_j)
\end{equation}
where we have represented the flow rate in the pipe using the standard
Poiseuille solution. 

We show in \apc{dim_analys} that the pressure loss in  the part of the
channel where the $s-\ell$ interfaces of the two grains do not face each
other can be neglected in comparison to the remaining terms. Therefore,
the pressure is constant in these locations, and decreases linearly from
$P_i$ to $P_j$ along $L_c$. This simple pressure profile in the liquid
channel can be integrated to obtain the forces exerted by the grains on
the liquid to yield
\begin{equation}
\ve{F}_a(P_i,P_j) =\left(
\begin{array}{c}
(P_j-P_i)h\\
P_i  L_{ia}+P_j L_{ja}\\
\end{array}
\right)
\label{fap}
\end{equation}
and
\begin{equation}
\ve{F}_b(P_i,P_j) =\left(
\begin{array}{c}
(P_j-P_i)h \\
 -P_i  L_{ib}-P_j  L_{jb}\\
 \end{array}
\right)
\label{fbp}
\end{equation}
Of course, an opposite force is exerted by the fluid on the
grains.

Next, we consider the individual effect of a grain displacement in the
normal direction, with $V^t_{O_a}=V^t_{O_b}=0$ and $P_i=P_j=0$
(\figc{elem}, case (b)). Equations (\ref{base}) and (\ref{flux_def}),
combined with a no-slip condition at the $s-\ell$ interfaces,
link the pressure gradient in the $X-$direction to the fluid
flow.
\begin{equation}
\Phi_{i\rightarrow j}(X) = -\frac{2}{3\mu} \ppd{p}{X} h^3
\end{equation}
With the help of \eqc{pompe}, we have then:
\begin{equation}\label{pressure_eqn}
\frac{2h^3}{3\mu} \pdd{p}{X}=V^n_b-V^n_a
\end{equation}
Integrating \eqc{pressure_eqn} twice in $X$, and considering the imposed
symmetry on the pressure gives
\begin{equation}
 p(V^n_a,V^n_b) = \frac{3 \mu \left( V^n_b-V^n_a\right)}{4 h^3}  \left[X^2-\left( \frac{L_c}{2 h}
 \right)^2\right]
 \label{pres1}
\end{equation}
This term represents the change in pressure induced by
the fluid flow required to compensate the channel expansion
($V^n_b>V^n_a$) or constriction ($V^n_b<V^n_a$). Equation (\ref{pres1})
can be integrated once more over the length of the channel to
obtain the forces exerted by the grains on the liquid:
\begin{equation}
\ve{F}_a(V^n_a,V^n_b) =\left(
\begin{array}{c}
0\\
 -\mu \left( {L_c}/{2 h} \right)^3 \left( V^n_b-V^n_a\right) \\
\end{array}
\right)
\label{favn}
\end{equation}
and
\begin{equation}
\ve{F}_b(V^n_a,V^n_b) =\left(
\begin{array}{c}
0\\
  \mu \left( {L_c}/{2 h} \right)^3 \left( V^n_b-V^n_a\right)\\
 \end{array}
\right)
\label{fbvn}
\end{equation}

The symmetry of the formulation implies that at the center of the
channel
\begin{equation}
\left.\phi_{i\rightarrow j}\right|_{X=0}(V^n_a,V^n_b) = 0
\end{equation}
As the fluid is incompressible, a flux balance is readily written at the
channel vertices as
\begin{equation}
\Phi_i(V^n_a,V^n_b) =  L_{ib} V^n_b - L_{ia} V^n_a
\end{equation}
\begin{equation}
\Phi_j(V^n_a,V^n_b) =  L_{jb} V^n_b - L_{ja} V^n_a
\end{equation}
The final case that we consider isolates the effect of tangential
displacement of the grains (\figc{elem}, case(c)). This situation
corresponds to pure shear of the liquid channel, and thus we have:
\begin{equation}
\ve{F}_a(V^t_a,V^t_b) =\left(
\begin{array}{c}
 -\mu (L_c/2h) (V^t_b-V^t_a)\\
0\\
\end{array}
\right)
\end{equation}
and
\begin{equation}
\ve{F}_b(V^t_a,V^t_b) =\left(
\begin{array}{c}
\mu (L_c/2h) (V^t_b-V^t_a)\\
  0\\
 \end{array}
\right)
\end{equation}
Note that we have neglected the shear forces in the portions of the channel
where the two grains do not face each other. Considering the flow relative
to the initial (reference) configuration, the average tangential
displacement of the grains induces a fluid flow at each vertex given by
\begin{equation}
\Phi_i(V^t_a,V^t_b) =   2h  \frac{V^t_b+V^t_a}{2}
\end{equation}
\begin{equation}
\Phi_j(V^t_a,V^t_b) =  - 2h  \frac{V^t_b+V^t_a}{2}
\end{equation}
If $V^t_a=-V^t_b$, there is no net flow in the channel\ie the
liquid experiences a perfectly symmetric shear stress.

For a more general situation in which all three of the phenomena just
discussed may occur, the various contributions to the fluid flow and to
the forces on the grains can be written as the sum of the individual
simple cases just described, with the result
\begin{equation}
\ve{F}_a =\left(
\begin{array}{c}
(P_j-P_i)h -\mu ({L_c}/{2h}) (V^t_b-V^t_a)\\
 -\mu \left( {L_c}/{2h} \right)^3 V^n+P_i  L_{ia}+P_j L_{ja}\\
\end{array}
\right)
\label{fa}
\end{equation}
\begin{equation}
\ve{F}_b =\left(
\begin{array}{c}
(P_j-P_i)h +\mu ({L_c}/{2h}) (V^t_b-V^t_a)\\
  \mu \left( {L_c}/{2h} \right)^3 V^n-P_i  L_{ib}-P_j  L_{jb}\\
 \end{array}
\right)
\label{fb}
\end{equation}
\begin{equation}
\Phi_i =  \left(\frac{2 {h}^3}{3 \mu L_c} + \frac{\pi r_p^4}{8\mu
D_{av}L_c}\right)  (P_i-P_j) + 2h  \frac{V^t_b+V^t_a}{2} + L_{ib}
V^n_b - L_{ia} V^n_a \label{phi}
\end{equation}
\begin{equation}
\Phi_j =  \left(\frac{2 {h}^3}{3 \mu L_c} + \frac{\pi r_p^4}{8\mu
D_{av}L_c}\right) (P_j-P_i) - 2h \frac{V^t_b+V^t_a}{2} + L_{jb}
V^n_b - L_{ja} V^n_a \label{phj}
\end{equation}
We now proceed to develop these expressions into a form suitable for
a finite element formulation.

\subsection{Elementary matrix}
\label{elem_mat}
In order to obtain a matrix form, we collect the primitive variables into a
vector $\ve{U}^T = \left(P_i, P_j, V^n_a, V^n_b, V^t_a, V^t_b\right)$, and
the associated flow rates and forces into a second vector
$\ve{W}^T = \left(\Phi_i,\Phi_j,F_a^n, F_b^n,F_a^t,F_b^t\right)$. The
superscript ``$T$'' indicates the transpose of the vector. Equations
(\ref{fa})-(\ref{phj}) can then be written in matrix form as
\begin{equation}\label{E}
\ve{W}=\left(
\begin{array}{c}
  \Phi_i \\ \Phi_j \\ F_a^n \\  F_b^n\\F_a^t \\  F_b^t \\
\end{array}
\right)=\ten{E}\ve{U}=\left(
\begin{array}{c c c c c c}
 +C_1 & -C_1 & - L_{ia} & + L_{ib} & +h & +h \\
-C_1 & +C_1 & - L_{ja} & + L_{jb} & -h & -h \\
+ L_{ia} &+ L_{ja} &+C_2&-C_2& 0 & 0\\
- L_{ib} &- L_{jb} &-C_2&+C_2& 0 & 0\\
-h&+h&0&0&+C_3&-C_3\\
-h&+h&0&0&-C_3&+C_3\\
\end{array}
\right) \left(
\begin{array}{c}
P_i \\ P_j \\ V^n_a \\  V^n_b \\V^t_a \\  V^t_b \\
\end{array}
\right)
\end{equation}
where
\begin{equation}
C_1= \left( \frac{2 {h}^3}{3 \mu L_c} + \frac{\pi {r_p}^4}{8 \mu
D_{av} L_c}  \right) \quad ;\quad C_2=\mu \left( \frac{L_c}{2 h}
\right)^3\quad ;\quad C_3= \mu \frac{L_c}{2h}
\end{equation}
Please note that, unlike the usual formulation of mechanical problems
using the finite element method, we have chosen to group together the
normal and tangential components of the forces and velocities acting on
grains $a$ and $b$. One should also keep in mind that the
component $V^n_a=V^n_{O_a}-\beta v^*$ (or $V^n_b=V^n_{O_b}+\beta v^*$)
includes both grain displacement and solidification shrinkage.

It is interesting to compute the power dissipation $\dot \Omega$
in the channel
\begin{equation}
\dot \Omega=\ve{V}_b \cdot \ve{F}_b+\ve{V}_a \cdot \ve{F}_a + P_i
\Phi_{i} + P_j \Phi_{j}
\end{equation}
or equivalently by:
\begin{equation}
\dot \Omega =\ve{U}^T \ten{E} \ve{U}
\end{equation}
Expanding \eqc{omega_dot_eqn} using \eqc{E}, we obtain
\begin{equation}\label{omega_dot_eqn}
\dot \Omega=\frac{2 {h}^3}{3 \mu L_c} (P_i-P_j)^2+\mu \left(
\frac{L_c}{2 h}\right)^3 (V^n_b-V^n_a)^2+\mu \frac{L_c}{2h}
(V^t_b-V^t_a)^2
\end{equation}
Thus, the element matrix of the system is  positive definite and
represents a quadratic form related to dissipation in the channel.

It is convenient to decompose the matrix $\ten{E}$ as the sum of its
symmetric and antisymmetric parts, $\ten{S}$ and $\ten{A}$, respectively
\begin{equation} \ten{S} =\left(
\begin{array}{c c c c c c}
 +C_1 & -C_1 & 0 & 0 & 0 & 0 \\
-C_1 & +C_1 & 0 & 0 & 0& 0 \\
0 &0 &+C_2&-C_2& 0 & 0\\
0 &0 &-C_2&+C_2& 0 & 0\\
0&0&0&0&+C_3&-C_3\\
0&0&0&0&-C_3&+C_3\\
\end{array}
\right)
\end{equation}
and
\begin{equation}
\ten{A}=\left(
\begin{array}{c c c c c c}
 0 & 0& - L_{ia} & + L_{ib} & +h & +h \\
0& 0 & - L_{ja} & + L_{jb} & -h & -h \\
+ L_{ia} &+ L_{ja} &0&0& 0 & 0\\
- L_{ib} &- L_{jb} &0&0& 0 & 0\\
-h&+h&0&0&0&0\\
-h&+h&0&0&0&0\\
\end{array}
\right)
\end{equation}
Inserting this decomposition into the power dissipation yields
\begin{equation}
\dot \Omega= \ve{U}^T (\ten{S}+\ten{A})
 \ve{U}= \ve{U}^T \ten{S} \ve{U}
\end{equation}
since $\ve{U}^T \ten{A} \ve{U} \equiv 0$. The antisymmetric part of
$\ten{E}$,  $\ten{A}$, represents coupling
between the pressure in the fluid channels and the displacement of
the grains. This term does not dissipate energy in our formulation.

This form of the matrix is consistent with the Onsager-Casimir theory of
transport phenomena. \cite{casimir} Consider the two sets of conjugate
quantities. One set is invariant to time reversal
($P_{i},P_j,\ve{F}_a,\ve{F}_b$), whereas the other set
($\Phi_i,\Phi_j,\ve{V}_a,\ve{V}_b$) changes sign with time reversal.  The
symmetric matrix $\ten{S}$ couples quantities that have different behavior
with respect to time reversal. The anti-symmetric matrix $\ten{A}$ couples
the quantities with the same behavior.  For example, $\Phi_i$ is related
to $P_{i}$,$P_j$ by $\ten{S}$ and is related to $\ve{V}_a$,$\ve{V}_b$ by
$\ten{A}$. It is interesting to note that this fundamental relation is
obtained by a simple integration of fluid flow equation \cite{sek-mat}.
The usual volume-averaged formulation with the same set of unknowns for
describing semi-solid materials (pressure in the liquid, velocity or
deformation rate in the solid) does not reveal the symmetries inherent in
the present model. 


\subsection{Global problem}
\label{sect:global}

\begin{figure}[htb]
\begin{center}
\includegraphics[height=2 in]{\COM/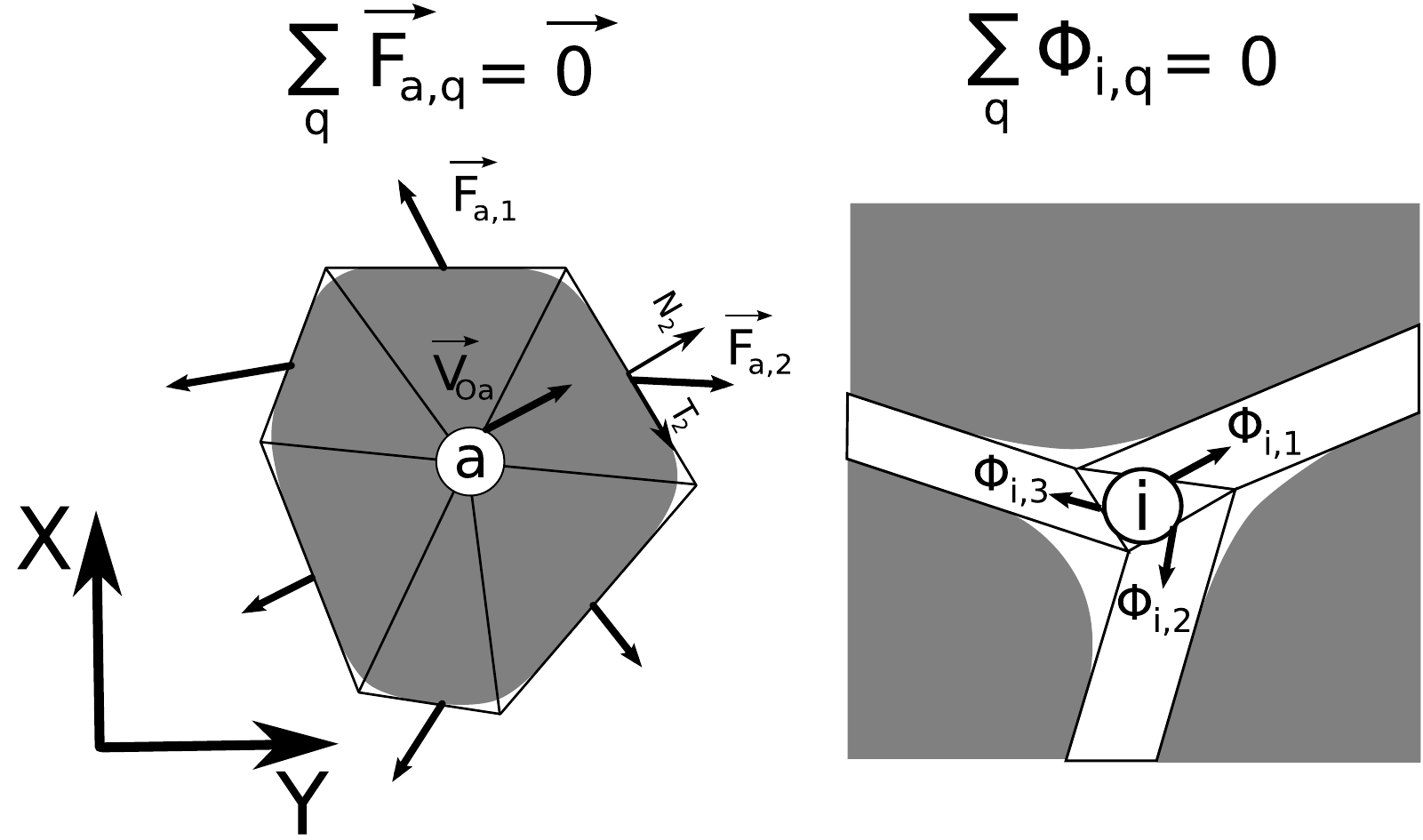}
\caption{Construction of the global problem from the elementary matrices:
The sum of forces on each grain is zero and the sum of fluxes flowing in
and out of each vertex is zero.} \label{masse}
\end{center}
\end{figure}

To assemble the global problem from the contributions of the individual
elements, both the unknown velocities at the grain centers and the resulting
forces are expressed in a global coordinate system $(X,Y)$ (see
\figc{masse}). One has for example:
\begin{equation}
\left(\begin{array}{c}F_a^X \\ F_a^Y\end{array}\right)
=\left(\begin{array}{c c}n_X & -n_Y\\n_Y & n_X\end{array}\right) 
\left(\begin{array}{c}F_a^n \\
F_a^t\end{array}\right)
\end{equation}
where $N_X$ and $N_Y$ are the components of the channel normal vector in
the global frame.  The pressures and fluxes for each channel in the global
coordinate system are obtained by the usual transformation rules\ie
\begin{equation}
\ve{W}'=\ten{Q}\ve{W}=\ten{Q}\ten{E}\ve{U}=
\ten{Q}\ten{E}\ten{Q}^T\ve{U}'=\ten{E}'\ve{U}'
\end{equation}
where $(\ve{W}')^T=(\Phi_i, \Phi_j, F_a^X,
F_b^X, F_a^Y,F_b^Y)$, $(\ve{U}')^T=(P_i, P_j, V_a^X,
V_b^X, V_a^Y,V_b^Y)$ and $\ten E'=\ten Q\ten E\ten Q^T$. The
transformation matrix $\ten {Q}$ is given by:
\begin{equation}
\ten Q= \left(
\begin{array}{c c c c c c}
1 & 0 & 0 & 0 & 0 & 0\\
0 & 1 & 0 &0 &0 &0\\
0 & 0 & n_X &0 &-n_Y &0\\
0 & 0 &0 &n_X &0 &-n_Y \\
0 & 0  &n_Y &0&n_X &0\\
0 & 0 &0  &n_Y &0 &n_X\\
\end{array}
\right)
\end{equation}
Since $V_a^n=V_{O_a}^n-\beta v^*$ and $V^n_b=V^n_{O_b}+\beta v^*$,
\eqc{E} in the $(X,Y)$ frame becomes
\begin{equation}\label{global_element}
\left(
\begin{array}{c}
  \Phi_i \\ \Phi_j \\ F^X_{a} \\F^X_{b} \\  F^Y_{a}  \\  F^Y_{b} \\
\end{array}
\right)= \ten Q \ten{E} \ten Q^T \left(
\begin{array}{c}
P_i \\ P_j \\ V^X_{O_a}  \\ V^X_{O_b} \\  V^Y_{O_a} \\  V^Y_{O_b} \\
\end{array}\right) + \ten Q\ten E\left(
\begin{array}{c}
0 \\ 0 \\ -\beta v^*  \\ +\beta v^* \\  0 \\ 0\\
\end{array}\right)
\end{equation}
The last term, which we call $\ve {B}'$, is associated
with solidification shrinkage and is known, coming from the external
solidification model.
The balance of mass and force is obtained by summing \eqc{global_element}
over all elements and setting the result to zero. This procedure is similar to the matrix assembly in the standard finite element method\ie each contribution associated with a given liquid channel with a local numbering (a, b, i, j) (\secc{elem_mat}) is added to the global matricial problem with a global numbering of all the grains and vertices.
The result is written in the compact form
\begin{equation}
\ten{E_{tot}}' \ve{U_{tot}}' = -\ve{B_{tot}}' \label{Global}
\end{equation}
where the vector of unknowns $\ve{U_{tot}}$ contains the velocities of the
$N_g$ grains $(V_{O_a}^X,V_{O_a}^Y)$ in the global frame and the pressures
$P_i$ at the $N_v$ vertices.

Integration points located at the external boundary of the global domain
are subject to boundary conditions. An imposed flux on a channel vertex
and an imposed force on a grain are Neumann-type boundary conditions in
this formulation. These are taken into account by adding the imposed
constraint into the global vector  $-\ve{B_{tot}}'$. Boundary conditions
specifying either the velocity of a grain or an imposed pressure at a
channel vertex are essential boundary conditions.   These are included in
the formulation using a penalty method, as in a standard Finite Element
Method.

Finally, the linear system of \eqc{Global} is solved with a standard LU decomposition.
 The velocities computed in the $\ve{U_{tot}}$ are used to update the grains positions at the next time step.


\subsection{Detection of contact}
\label{contact}

The present model is intended to be used for the study of hot cracking\ie
for $g_s \gtrsim 0.9$, so that the width of the liquid channels is very
small compared to the grain size, and the displacement of the grains
will be very limited. It is important in the implementation of the model
to detect and account for contact between grains. Equation
(\ref{vol_flow_rate}) shows that the pressure drop in a channel tends to
infinity as $(1/h)^3$ as $h \rightarrow 0$, and this  is sufficient to
prevent the interpenetration of the grains in the simulation, as long as 
the time step is sufficiently small. In some highly constrained
situations, we found it necessary to implement a dynamic time step
refinement procedure in the code in order to prevent grain
interpenetration.


It is more problematic to handle the changes of the environment of
the grains\ie modifications of their nearest neighbors. This
typically occurs after large displacements of the grains, especially
for channels that are short to begin with. There are a few
additional situations where the grains can interpenetrate (see
\cite{Vernede2007} for further details). These cases are handled by
revising the list of the first neighbors after each time step.
\cite{Ferrez2001}. In the example problems presented in the next
section, the volume fraction of solid is high, and the system is thus
sufficiently constrained that this phenomenon does not occur. 



\section{Results}
\subsection{Boundary conditions}

\begin{figure}[htb]
\begin{center}
\subfigure[Ideal feeding\label{cl_nonB}]
   {\includegraphics[width=0.4\textwidth]{\COM/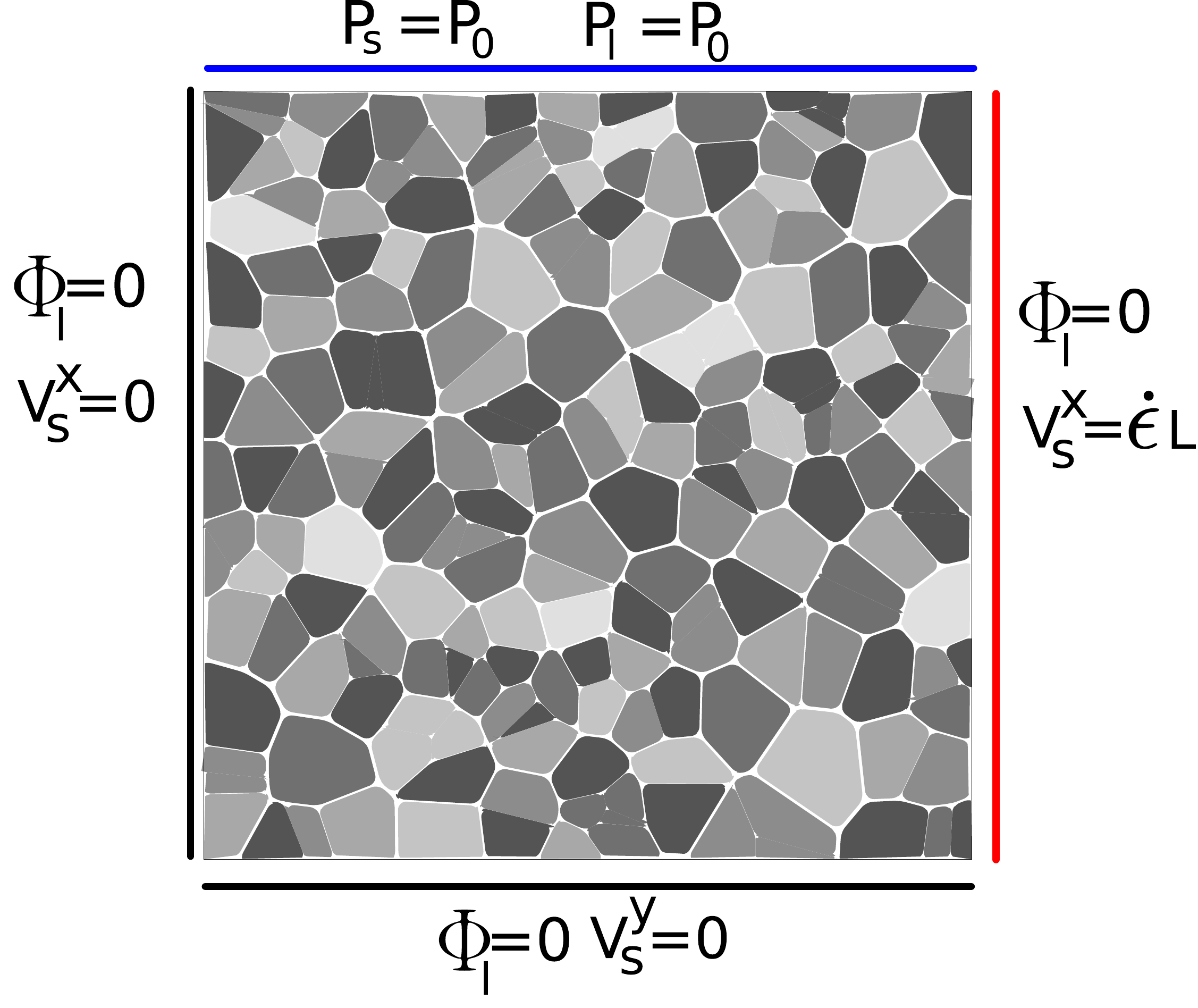}}
\subfigure[No feeding \label{cl_B}]{\includegraphics
[width=0.4\textwidth]{\COM/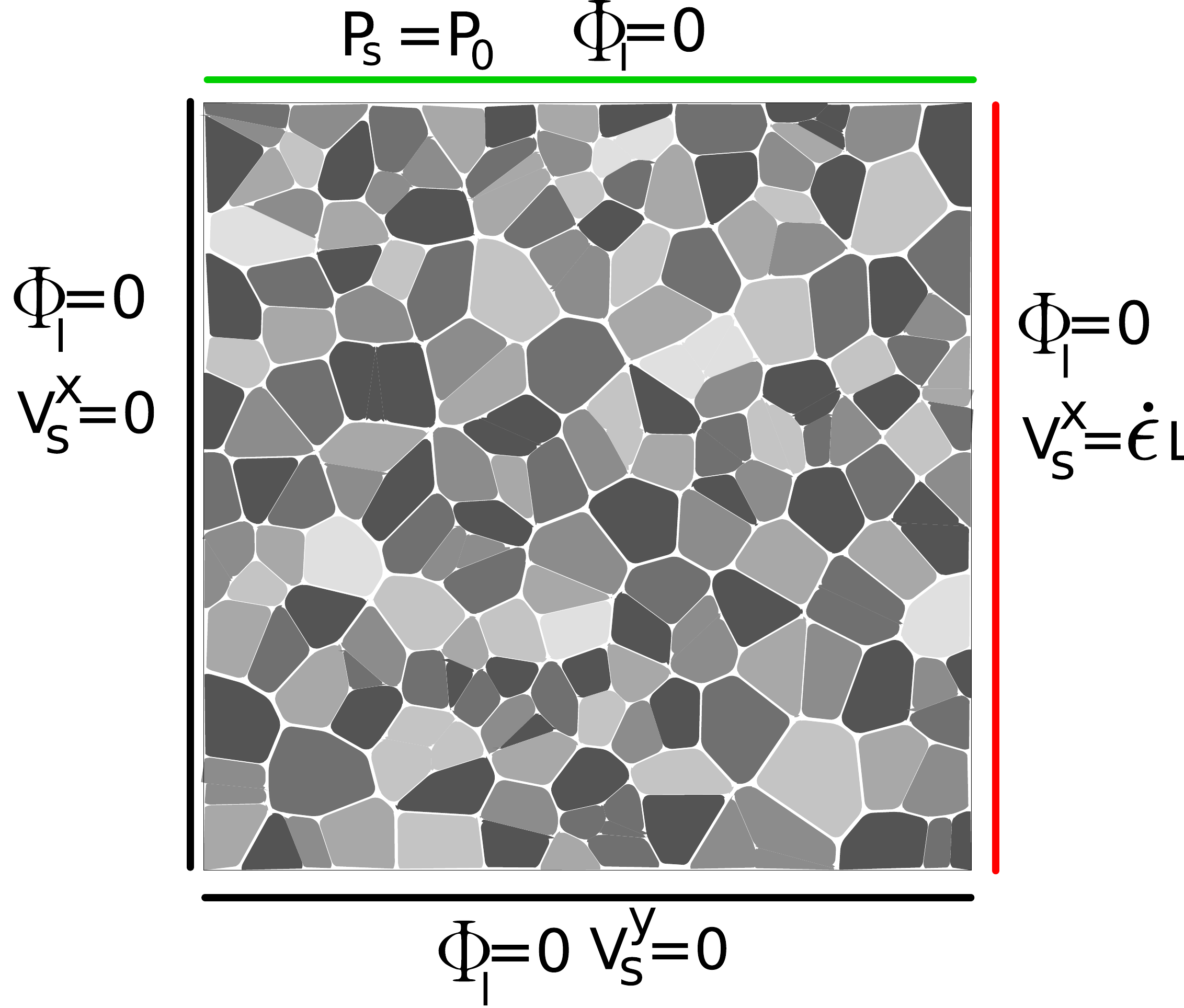}} 
\caption{Boundary conditions
for two traction tests along the $X$ axis: \subref{cl_nonB} Ideal
feeding from the upper boundary; \subref{cl_B} No feeding. The
grains or clusters of grains are colored with various grey levels.}
\label{cl}
\end{center}
\end{figure}

To use the model to investigate the mechanical behavior of the mushy zone,
two test cases were analyzed where a tensile load was applied to a small
ensemble of grains comprising a square volume element of edge $L_V$.
Boundary conditions for the two cases are shown in \figc{cl}. Since two
phases are present, boundary conditions are needed for either the pressure
or the flow rate for the liquid channels, and for either the velocity or
the applied forces on the solid grains.

The two cases differ only in the boundary conditions applied on the top
surface. In the first case (\figc{cl_nonB}), a pressure $P_0$ is imposed
on both the liquid and the solid, corresponding to an average hydrostatic
pressure on the system.  \footnote{To represent the hydrostatic pressure
on the solid, a force is imposed on each grain at the boundary. This force
is oriented along the normal to the boundary and is equal to $-P_0L_b$
where $L_b$ is the length of the external boundary of the grain.} The
fluid flow is free, thus allowing feeding from the upper boundary. We
refer to this case as ``ideal feeding'' because it simulates a sample in
contact with a liquid reservoir (a feeder) at pressure $P_0$. In the
second case, we also impose a pressure $P_0$ on the upper surface of the solid grains but the
fluid flow is set to zero. This simulates a situation where feeding is
impossible.

The remaining boundary conditions are the same for both cases.  The fluid
flow and the horizontal $X$-component of the grain velocity are zero on
the left boundary, while the $Y$-component of the grain velocity is
free\ie no forces along the vertical axis. This set of boundary conditions
is equivalent to a symmetry plane. On the bottom boundary, the flux and
the $Y$-component of the velocity are zero, while the $X$-component is
free. On the right boundary, a velocity $\dot\varepsilon L_V$ is imposed
on the solid in the $X$-direction to study the effect of an imposed strain
rate $\dot\varepsilon$. The fluid flow is zero on the right
boundary. \footnote{Note that for boundary conditions, the liquid flux is
considered in the frame of the solid grains. In the laboratory frame, a
fluid flux is observed due to the advection of the solid.}

We note that these boundary conditions are similar to those of the model derived by Lahaie
and Bouchard for a regular arrangement of hexagonal grains \cite{Lahaie}.
In our numerical calculations, the solidification of the system is
calculated first, and then the mechanics of the mushy zone is computed
without allowing any further solidification\ie at fixed solid fraction.
This implies that $g_s$ is fixed, so that there is no
solidification shrinkage.

\subsection{Tension tests}

\begin{figure}[htb]
\begin{center}
\subfigure[\label{lowfs_duct}]{\includegraphics[width=0.49\textwidth]{\COM/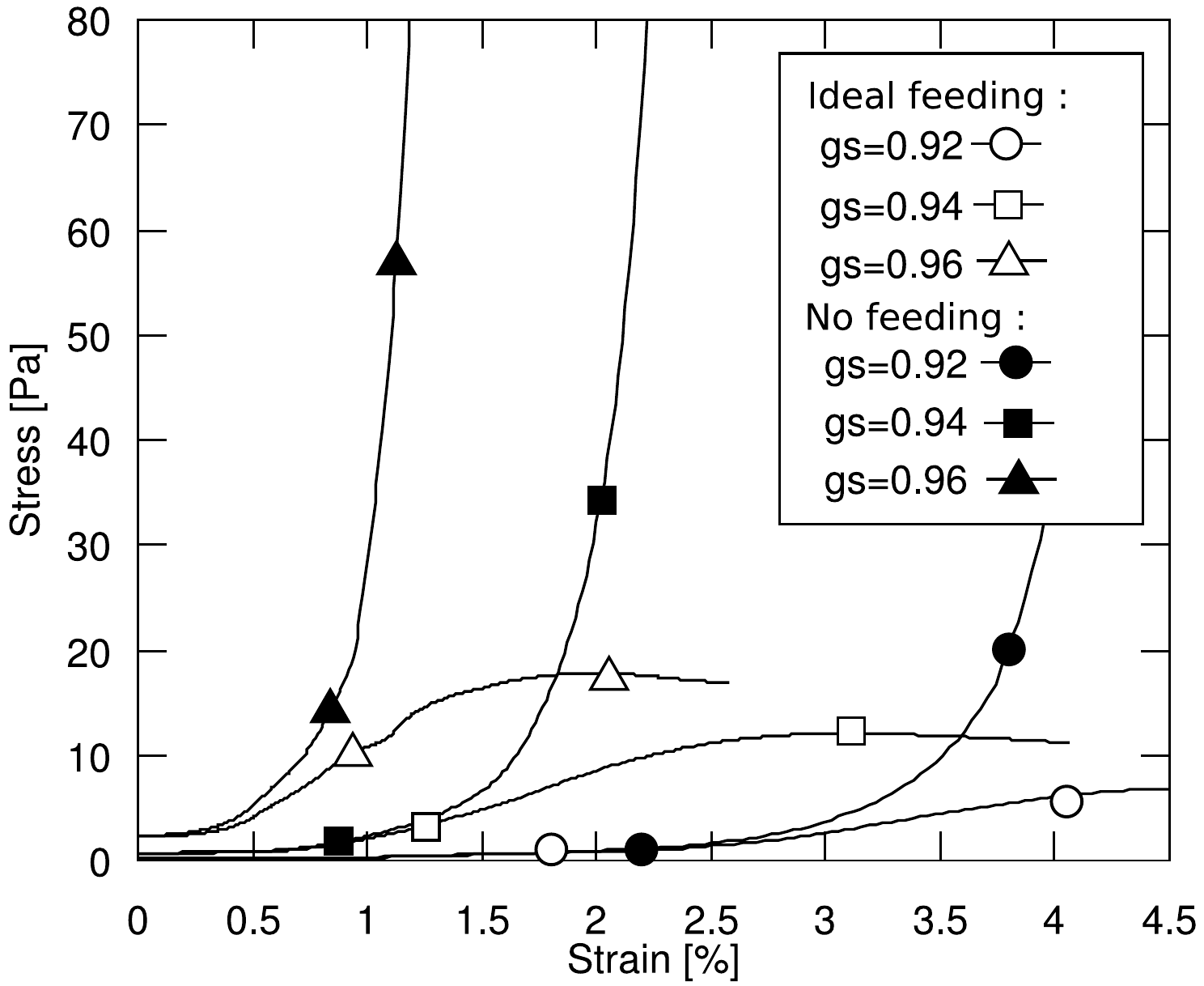}}
\subfigure[\label{highgs_duct}]{\includegraphics
[width=0.49\textwidth]{\COM/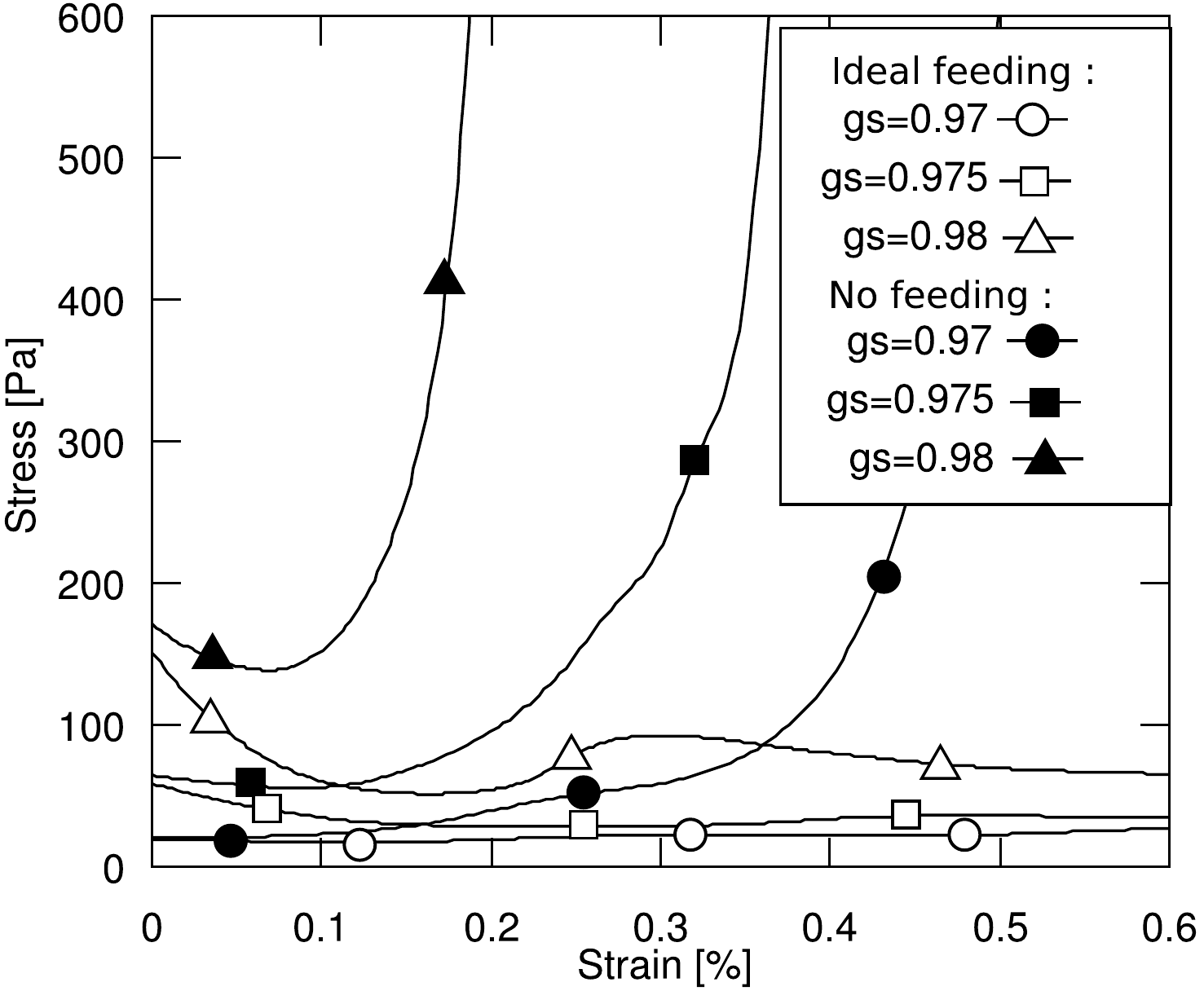}} \caption{Stress as a
function of strain for a square domain strained at a rate $\dot
\varepsilon=4\times 10^{-3}$ \fo{s^{-1}}. Open and filled symbols
correspond to ideal and no feeding, respectively. \subref{lowfs_duct}
$g_s<0.97$, \subref{highgs_duct} $g_s>0.97$ .}
\label{high_low_gs}
\end{center}
\end{figure}

Figure \ref{lowfs_duct} shows the stress-strain curves for a sample
solidified at constant $\dot{T} =-1$ \fo{K~s^{-1}}, up to three
volume fractions of solid ($g_s=0.92$, $g_s=0.94$, and $g_s=0.96$).
Each sample was then strained along the $X$-direction at a rate
$\dot \varepsilon=4\times 10^{-3}$ \fo{s^{-1}}. The results for the two
tests, ideal and no feeding, are shown with open and filled symbols,
respectively. As can be seen, the two tests give the same
stress-strain response at strains up to about 2.5\% for $g_s=0.92$,
1.2\% for $g_s=0.94$ and 0.3\% for $g_s=0.96$. Beyond these strains,
the stress increases abruptly when feeding is not allowed, whereas
it remains low and even decreases past a maximum in the case of
ideal feeding.

\begin{figure}[htbp]
\begin{center}
\includegraphics [width=\textwidth]{\COM/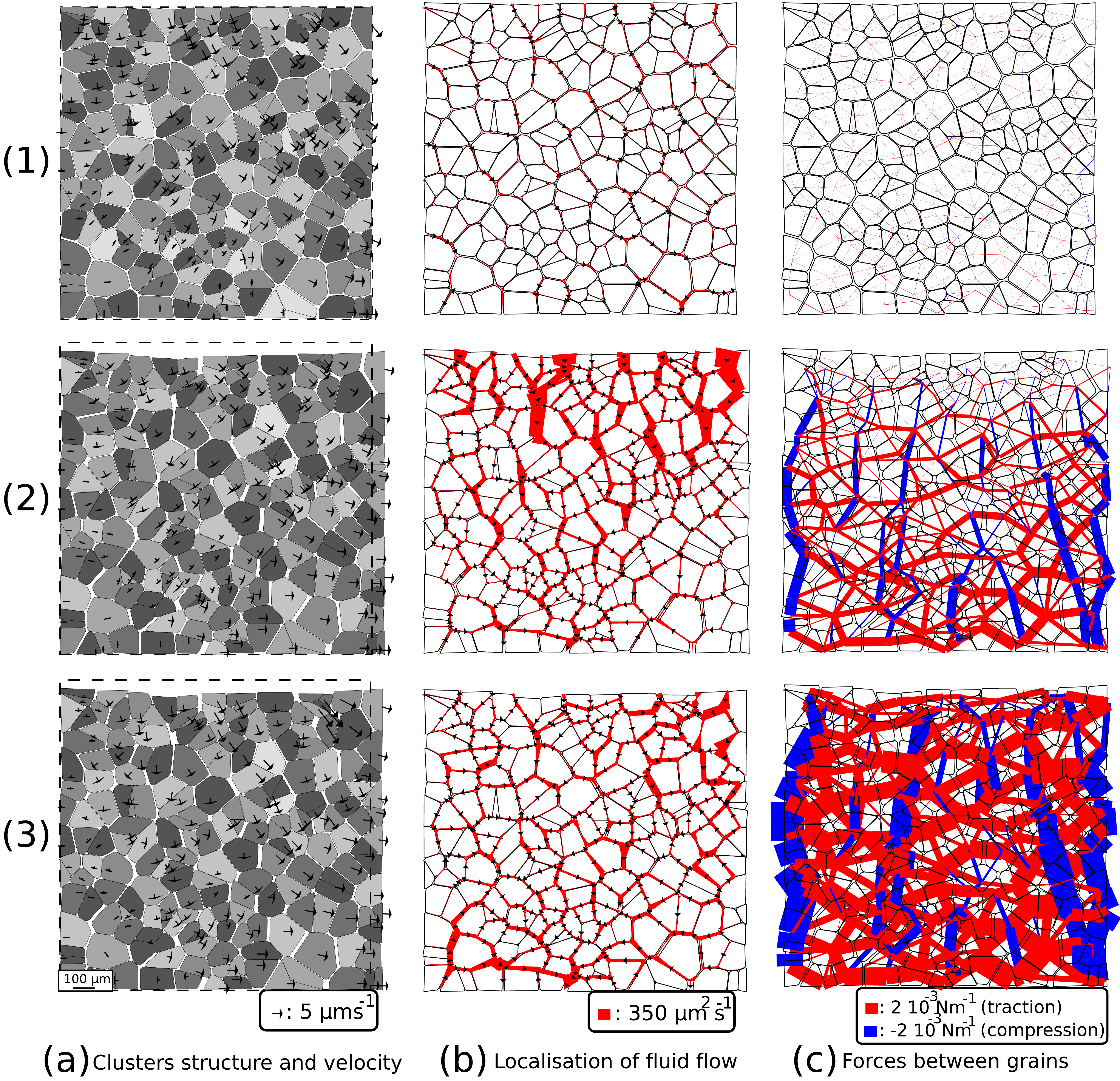}
\caption{Isothermal mushy zone with $g_s=0.92$ and deformed at a
strain rate $\dot \varepsilon=4\times 10^{-3}$ \fo{s^{-1}} along the
horizontal $X$-direction: (1) Mushy zone at 0\% strain; (2) Mushy
zone at 4\% strain with ideal feeding from the upper face; (3) Mushy
zone at 4\% strain without feeding from the upper face. In (a), the
grains (or grain clusters) are shown with various grey levels
together with their velocity represented with small arrows. In (b),
the flow in the channels is represented with a line of thickness
proportional to the intensity and a triangle indicating the
direction. In (c), the forces between the grains are represented
with lines of variable thickness proportional to the modulus. A grey hue (red) corresponds to traction and a dark hue (blue) to compression.}
\label{Vblock_gs092e04}
\end{center}
\end{figure}

These results can be understood by looking at the local deformation
mechanisms shown in \figc{Vblock_gs092e04}, corresponding to the
isothermal mushy zone strained at $g_s=0.92$. This domain contains
200 grains having an average diameter of 100 \fo{\mu m}. Note
that the computation of a  traction test simulation on such a mushy
zone takes about 30 s on a personal computer with a 2 GHz Intel Core
Duo processor. Figure \ref{Vblock_gs092e04}(1) shows the grain
structure with the liquid channels at the onset of deformation\ie
for $\varepsilon = 0$. The grains are identified by assigning
various gray levels, and the velocity of each grain is displayed
with small arrows in \figc{Vblock_gs092e04}(1a) (the scale at the
bottom of the figure gives the modulus of the velocity). When two
grains  establish a solid bond\ie the width of the corresponding
channel goes to zero, they are shaded with the same gray scale,
making it easy to recognize the formation of grain clusters as
solidification progresses. \cite{second} At this relatively low
volume fraction of solid ($g_s=0.92$), only a few such clusters of
grains have formed.

In \figc{Vblock_gs092e04}(1b), the fluid flow in each channel is
represented by lines whose width is proportional to their magnitude, the
direction being indicated by a black triangle. The scale used for this
flow representation is again shown at the bottom of the figure. As two
grains get closer together, they squeeze the liquid out of the channel,
and on the other hand, as they move away from each other, liquid is pumped
into the channel. In both cases, the width of the corresponding line
varies along the channel length. The flow can also be important in
channels of fixed width that feed other regions of the mushy zone. As can
be seen in \figc{Vblock_gs092e04}(1b), deformation of the mush is
accommodated by fluid flow, but these flows remain small and localized.
Long-range feeding of the mush is not necessary at low strain, and
therefore both cases give the same behavior of the mush (see
\figc{lowfs_duct}).

The stresses in the sample are also very low at low strain, as
indicated in \figc{Vblock_gs092e04}(1c). The interaction forces
between the grains via the liquid channel are represented by a line
connecting nucleation centers, whose width is proportional to the
magnitude of the force (scale shown at the bottom), and whose gray
scale level (or colour) indicates whether the force is tensile or compressive.
At low volume fraction of solid and strain, the forces are of course
very low and the corresponding lines are barely visible in
\figc{Vblock_gs092e04}(1c).

Figures \ref{Vblock_gs092e04}(2) shows the same mushy zone after 4\%
deformation for the ideal feeding case, and
Fig.~\ref{Vblock_gs092e04}(3) shows the case where feeding is
prohibited.  The overall deformation is indicated in (a), the initial
volume element is drawn as a dashed-line square. It is important to note
that deformation is  localized to a few channels, roughly oriented
normally to the tensile direction.  The fluid tends to flow \emph{from}
channels oriented in the direction of the stress \emph{to} channels
oriented perpendicular to the stress. The grains tend to be pulled inward
along the vertical $Y$-direction as we try to pull the mushy zone in the
$X$-direction. This is also reflected by the forces shown in
\figc{Vblock_gs092e04}(2c) and (3c), which essentially correspond to
traction along the horizontal $X$-direction and to compression along the
$Y$-direction.

It is also interesting to note that fluid flow is much more
important in the case of ideal feeding (\figc{Vblock_gs092e04}(2b))
compared to the case of no feeding (\figc{Vblock_gs092e04}(3b)),
even though the imposed strain rate is the same. This shows that
redistribution of fluid occurs over larger distances with the
accumulation of deformation. Fluid flow from the upper boundary is
clearly visible in \figc{Vblock_gs092e04}(2b), and this flow relaxes
the stresses in the upper part of the sample
(\figc{Vblock_gs092e04}(2c)).

If the mush cannot be fed from the upper boundary
(\figc{Vblock_gs092e04}(3)), redistribution of the fluid channels and of
the grains also occurs as deformation increases. However, since no feeding
from the top surface is allowed, the fluid follows a more difficult path,
and the stresses in the samples are higher. Note that, as no liquid flow
is allowed on any boundary for this case and the solid grains are rigid,
the total volume of the specimen (grains + liquid) is constant.

In summary, at low strain, deformation is accommodated by local
redistribution of the liquid. This deformation is localized in
liquid channels oriented roughly perpendicular to the stress direction.
As deformation increases, more channels get closed and
fluid redistribution occurs at a larger scale. At that point, the
ability to feed the mush from some liquid reservoir becomes
important.

\subsection{Transition in the feeding mechanism}

As the volume fraction of solid in the volume element increases, the
difference between the ideal-feeding and the no-feeding cases occurs at
lower strain (\figc{lowfs_duct}). The reason for this is fairly obvious,
as the width of the liquid channels decreases with increasing $g_s$.

\begin{figure}[htb]
\begin{center}
\includegraphics [width=0.8\textwidth]{\COM/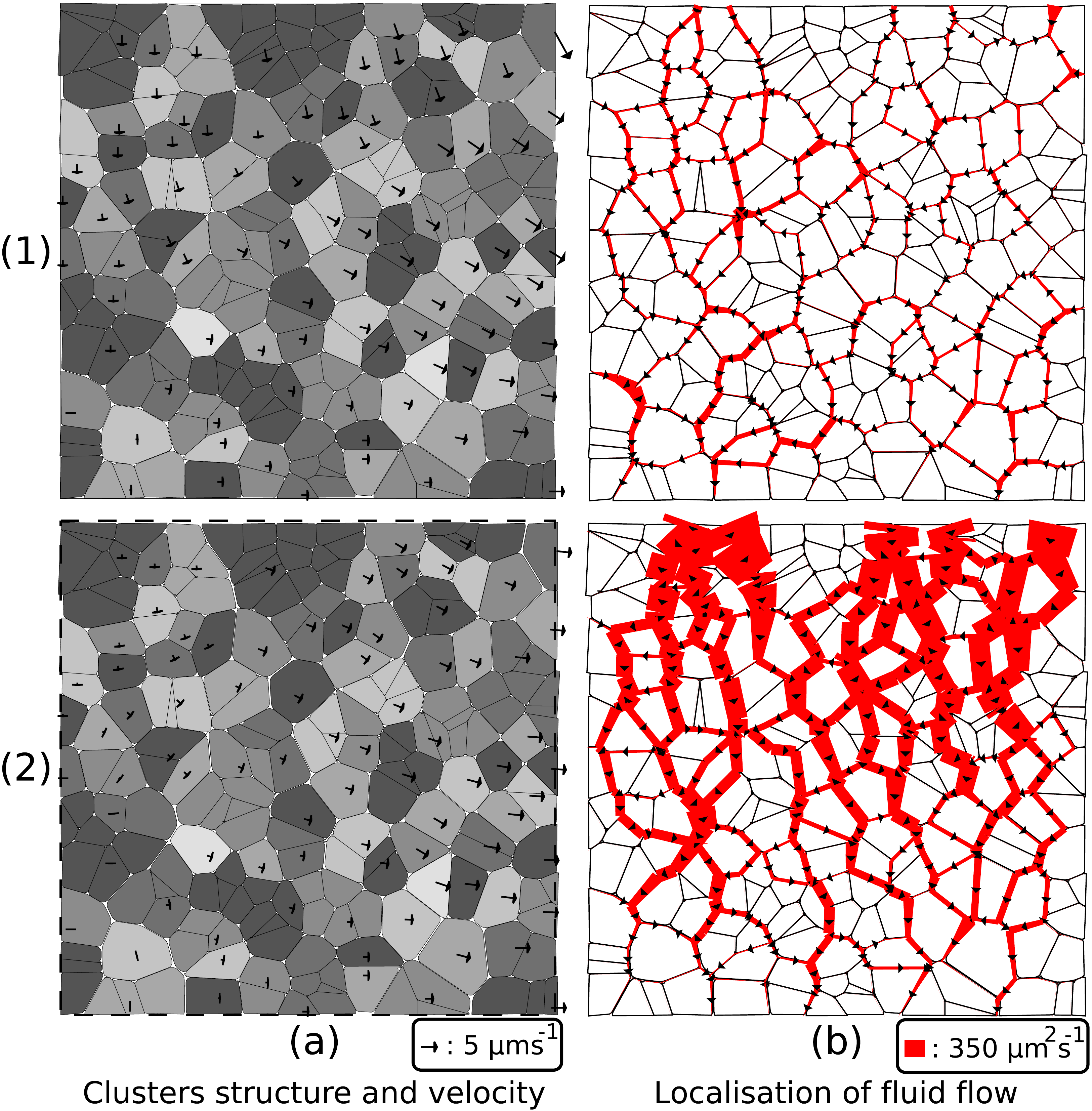}
\caption{Isothermal mushy zone with $g_s=0.975$ strained with $\dot
\varepsilon=4\times 10^{-3}$ \fo{s^{-1}}. Feeding from the upper face is
allowed: $\varepsilon = 0$ (1), $\varepsilon = 0.05\%$ (2). In (a), the
grains and grain clusters are shown with various grey levels and their
velocity is indicated with small arrows. In (b), the flow in the channels
is displayed with lines of thickness proportional to the intensity and
triangles indicating the direction.}
\label{V_gs0975e0}
\end{center}
\end{figure}

However, for $g_s > 0.97$ (\figc{highgs_duct}), the behavior of the two
cases becomes different even at the limit $\varepsilon \rightarrow 0$. To
understand this new response, consider Fig. \ref{V_gs0975e0}, which shows
a mushy zone with $g_s=0.975$ deformed at the same strain rate $\dot
\varepsilon=4\times 10^{-3}$ \fo{s^{-1}}. At the onset of deformation
($\varepsilon =0$, \figc{V_gs0975e0}(1a)), it can be seen that the solid
grains start to form significant numbers of grain clusters. Moreover,
\figc{V_gs0975e0}(1b) shows that there is a significant redistribution of
fluid, even though the deformation of the mush is zero at that stage. The
fluid is located predominantly at the boundaries of the clusters which
move as a single larger grain. It is interesting to compare this figure
with \figc{Vblock_gs092e04}(1b), which represents a mushy zone under the
same straining and feeding conditions, but at a lower solid fraction.

The same mushy zone is represented in \figc{V_gs0975e0}(2) at 0.05\%
deformation \ie after 0.125 s. The very small deformation is localized to just a few
channels that are located at the edges of the clusters and oriented normal
to the tensile axis. Local redistribution of the liquid is difficult and
most of the fluid to accommodate deformation is brought from the upper
boundary even at this low strain level. Therefore, above $g_s=0.97$, the
deformation mechanism due to local fluid redistribution is no longer
possible, due to the presence of relatively large clusters and thinner
liquid channels. It is interesting to note that the volume fraction of
solid at which the mechanical behavior of the mush changes corresponds
precisely to the value $g_{s,1\%ilc}$ at which 1\% of the liquid channels
become isolated and the permeability of the mush deviates from the
Kozeny-Carman relationship.\cite{second}

The mechanical model we have presented shows that this solid
fraction also corresponds to the point where accommodation of
deformation by fluid flow becomes extremely difficult. Therefore,
this point can be associated with the ductility minimum point
observed experimentally \cite{rev_crique,Magnin}, also called
coalescence solid fraction by some authors \cite{ludwig_model}. We
find this point at a value $g_s=0.97$, whereas the experiments
performed on inoculated globular microstructures give instead a
value $g_s=0.95$ \cite{ludwig_model}. This difference is probably
due to the 2D nature of our model, which tends to underestimate the
size of the last liquid films (segments in a 2D Voronoi model
instead of polyhedral surfaces in 3D).

\subsection{Discussion}

The effect of external loads on the response in the model can be
deduced easily from the linearity of the matrix $\ten{E}$ (\eqc{E}).
In particular, an increase of the metallostatic  pressure $P_0$
applied on the upper surface of \figc{cl} simply shifts the stress
response of the solid network.\footnote{From a numerical point of
view, it is clear that zero grain displacement and a uniform
pressure is a solution of the problem. Because the system is linear,
this implies that any solution can be shifted  by a uniform amount
of the pressure without affecting the displacement of the grains.}
This response corresponds to the Terzaghi effective stress\ie the
behavior of the mush depends only on the difference between the
applied stress and the hydrostatic pressure \cite{Terzaghi1943}.
Similarly, the stress response of the mushy zone varies linearly
with the strain rate.

On the other hand, the evolution of the mushy zone with strain
requires a numerical calculation. The stress-strain curves in
\figc{high_low_gs} can be compared qualitatively with experimental
data (see {\it e.g.} \cite{rev_crique,ludwig_model,Vernede2007}).
For solid fractions lower than $g_{s,1\%ilc}$ (\figc{lowfs_duct}),
the strain at which the stress increases abruptly corresponds well
to the experimental strains at fracture for the corresponding value
of $g_s$. However, the shape of the stress-strain curve and the
magnitude of the stress are clearly different from the experimental
values. At higher solid fractions, the strain at which the stress
increases abruptly no longer correlates well with the experimental
observations.

We have not considered solid deformation in the model. For
solid fraction lower than the ductility minimum
($g_s<g_{s,1\%ilc}$), solid deformation certainly plays a role in
the overall behavior of the mush, but grain displacement is the
dominant deformation mechanism. The model reproduces the strain at
which interlocking of the grains occurs, even though rotations were not
considered, but it cannot reproduce the shape of the stress-strain
curve. At higher solid fractions ($g_s>g_{s,1\%ilc}$), solid
deformation is clearly the dominant mechanism. The ductility
increase observed experimentally at high $g_s$-value corresponds to
the strength increase of the solid network due to its progressive
percolation.

This 2D model demonstrates a transition in the mechanical behaviour of the mush as the solid fraction increases. We have included the 3D aspects of that transition in a simplified way. The details of this transition requires a fully 3D model, which we will describe in a forthcoming publication.

 \subsection{First consequences on hot tearing criteria}

Hot tearing criteria based on the feeding ability of the mush
\cite{rdg,MHamdi2006,Grandfield2005} predict the hot cracking sensitivity
(HCS) of alloys  fairly well. Such criteria are based on a critical strain
rate, which seems to be particularly applicable in processes such as DC
casting, where thermally-induced stresses are essentially perpendicular to
the thermal gradient.\cite{Suyitno,Mathier2007}  However, in tensile test
experiments where the applied stress is parallel to the thermal gradient,
hot cracking is found to be largely independent of the stain
rate.\cite{Vernede2007,Mathier2007} As mentioned in the previous section,
the response of the present model is at first sight proportional to the
strain rate. However, the granular nature of the model makes it strongly
strain-dependent.

\begin{figure}[htb]
\begin{center}
\includegraphics[height=1.5 in]{\COM/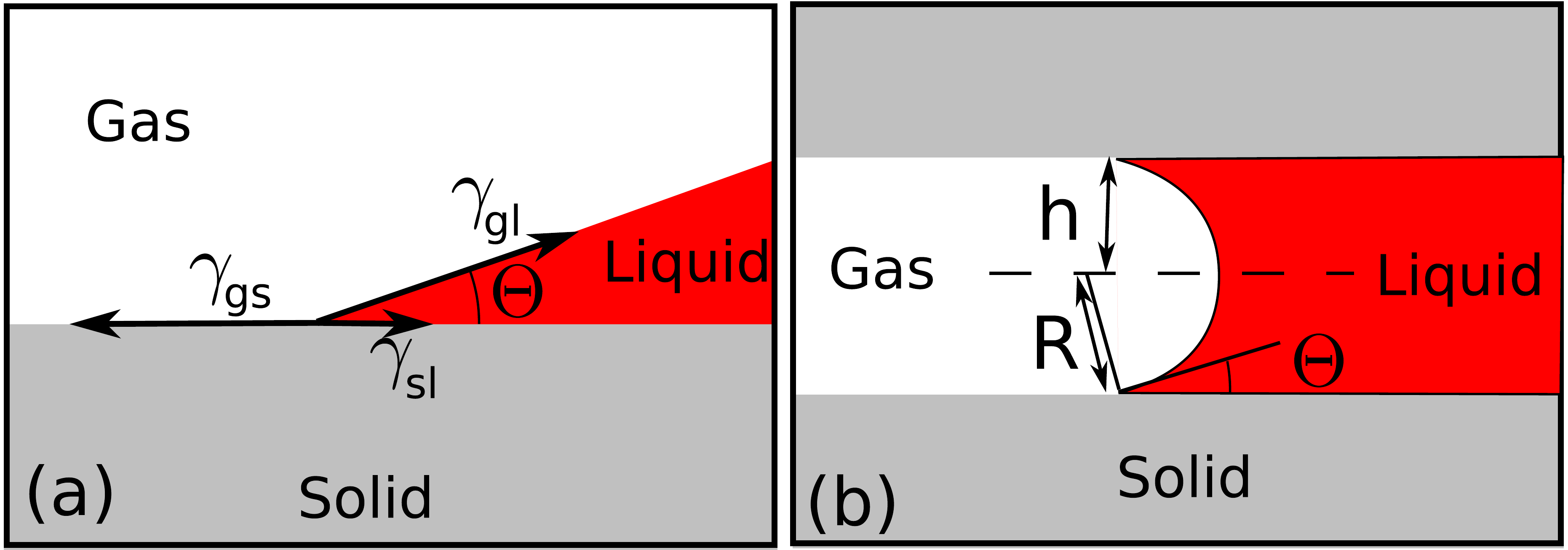}
\caption{(a) Equilibrium of forces at triple junction. (b) Shape of
the meniscus between two grains.}\label{triple}
\end{center}
\end{figure}

The criterion for hot tear nucleation is usually written in the form
\begin{equation}
p_l < p_c
\end{equation}
where $p_l$ is the local pressure in the liquid and $p_c$ represents the
cavitation pressure at which a pore nucleates. As a first approximation,
this cavitation pressure can be estimated as the overpressure required to
overcome capillary forces at the liquid-pore interface. Since pores will
form at high solid fraction\ie in very narrow liquid channels, their
radius of curvature is dictated by the width of the channel, as
illustrated in \figc{triple}. At the triple junction between the liquid,
solid and pore, the equilibrium condition is given by:
\begin{equation}
\gamma_{gl} \cos \Theta = \gamma_{gs}-\gamma_{sl} \label{eq_3p}
\end{equation}
where $\gamma_{sl}$, $\gamma_{gl}$ and $\gamma_{gs}$ are the interfacial
energy between solid and liquid, liquid and pore, and solid and
pore, respectively, and $\Theta$ is the dihedral angle. Therefore,
the radius of the pore $R$ is given by:
\begin{equation}
R=\frac{h}{\cos \Theta}
\end{equation}
where $h$ is the half width of the liquid channel (see \figc{triple}
(b)). Therefore
\begin{equation}
p_c = \Delta p_\gamma= -\frac{\gamma_{gl}}{R} =  -\frac{
\cos\Theta~\gamma_{gl}}{h}
 =  -\frac{\gamma_{gs}-\gamma_{sl}}{h}=  -\frac{I}{h}
\end{equation}
where $I$ is called the ``impregnation factor.''  Since liquid metals wet
very well their own solid\ie $\gamma_{sl}\ll\gamma_{gl}$, the value of $I$
is typically  close to $\gamma_{gl}$\eg 1 \fo{J~m^{-2}} for Al.
%

\begin{figure}
\begin{center}
\includegraphics [width=0.5\textwidth]{\COM/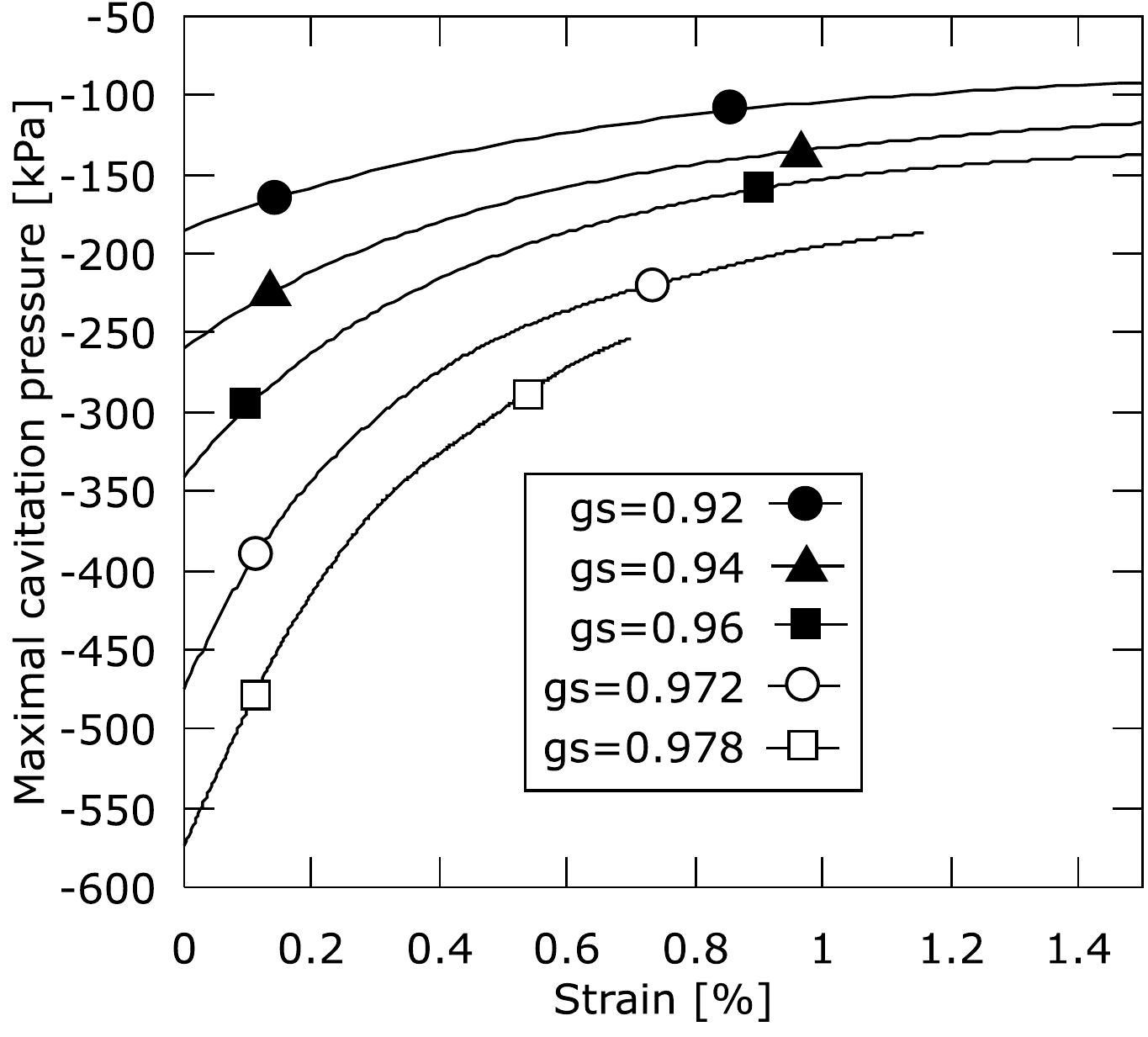}
\caption{Maximum value of the cavitation pressure in the mushy zone
as a function of strain for an impregnation factor $I$ of 1
\fo{Jm^{-2}}.} \label{Pc}
\end{center}
\end{figure}

Using such relationships, the maximal cavitation pressure  $p_{c,max}$ is
plotted in \figc{Pc} as a function of deformation for isothermal mushy
zones with various solid fractions deformed at a strain rate $\dot
\varepsilon=4\times 10^{-3}$ \fo{s^{-1}}. The pressure corresponds to the
pressure necessary for cavitation of a pore in the widest channel of the
mush, $h_{max}$. This channel width decreases with increasing $g_s$ and
considerably increases with strain, in particular for high solid fraction
samples. Note that this evolution is largely independent of the other
parameters.

\begin{figure}[htb]
\begin{center}
\includegraphics [width=\textwidth]{\COM/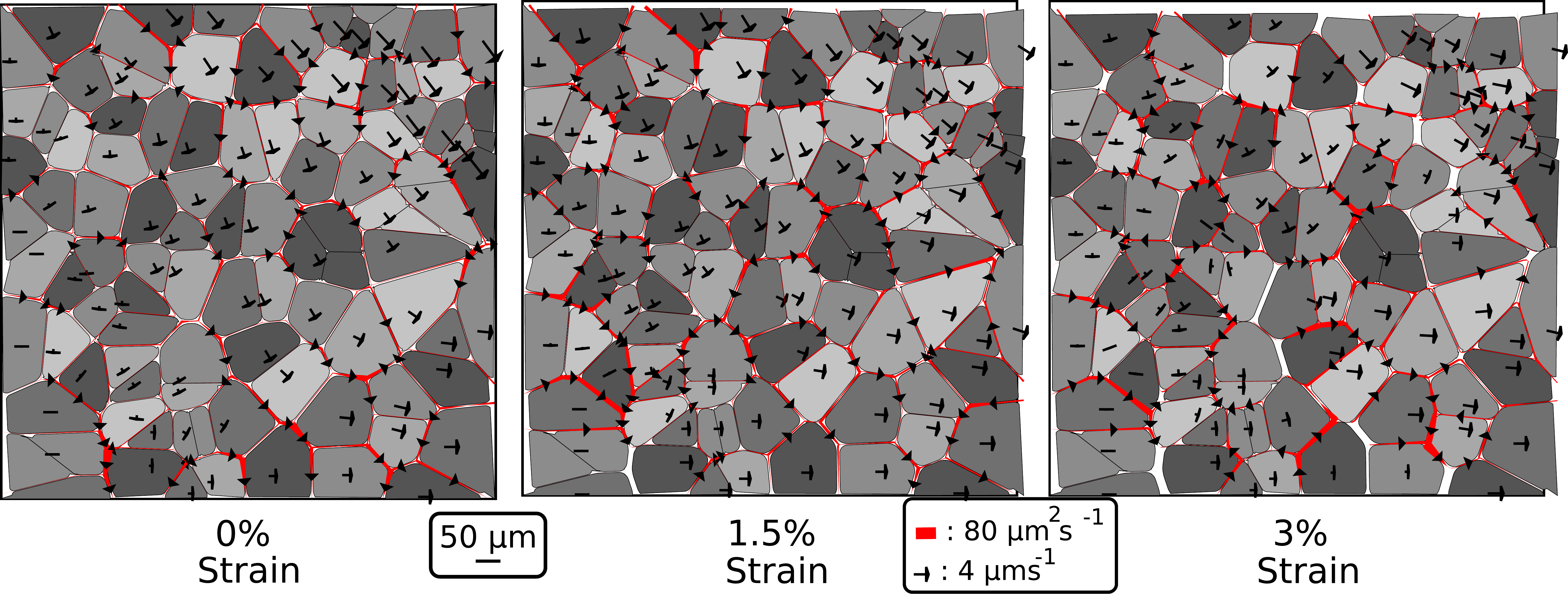}
\caption{Mushy zone at various strain levels with $g_s=0.92$ and
$\dot \varepsilon=4\times 10^{-3}$~\fo{s^{-1}}. Feeding from the upper face
is not allowed. Grain velocity (black arrows) and fluid flow (grey/red lines) are represented
 on the same picture. The channels on which feeding is not represented
  (white channels) correspond to those where a pore has nucleated.}
\label{cavitation}
\end{center}
\end{figure}

The variations of $p_c$ due to strain are therefore more important than
those induced by the strain rate. This phenomenon can explain the apparent
insensitivity of hot cracking to strain rate, at least for tensile test
experiments. This idea is further illustrated in \figc{cavitation} where a
mushy zone is represented at various strain levels. In this test, the
strain rate $\dot \varepsilon$ is equal to $4\times 10^{-3}$ \fo{s^{-1}},
$g_s=0.92$ and feeding from the upper boundary is not allowed. The
localization of the fluid flow and the grain velocity are represented on
the same picture. Channels in which feeding is not represented (white
channels), correspond to those in which a pore has nucleated. In order to
reach depressions capable of producing a pore by cavitation, the pressure at
the upper boundary was fixed to $P_0=-120$ kPa. The impregnation factor
$I$ in these simulations was fixed to 1 \fo{J~m^{-2}}.

Figure \ref{cavitation} clearly shows the nucleation of pores in the
largest channels, where deformation has been localized. Note that
this approach allows an estimation of the appearance of damage in
the mushy zone, but cannot model fracture, which  would require
explicit modeling of the deformation of the solid grains.

\section{Conclusion}

This paper has presented the development of a 2D granular model for
the mechanical behavior of mushy zones that brings valuable insight
into the interrelations between intergranular flow and grain
movement. The model is based on conservation of mass and force at
the level of the intergranular region. It produces naturally a form
that is consistent with non-equilibrium thermodynamics. The whole
mechanical problem can thus be expressed as a minimization of
dissipation. Therefore, the results of the present simulations are
well suited for continuum models based on a dissipation potential. 
\cite{ludwig_model}

The model accounts directly for the random nature of grain nucleation, and
for the progressive formation of grain clusters during solidification.
\cite{second} In addition, the granular nature of the material is modeled,
which leads to further localization of deformation and feeding (besides
the localization due to grain clusters). Both of these important phenomena
cannot be resolved using typical volume-averaged formulations that smear
out the details of the grain structure.

The formulation of the model requires only a few integration points for
each grain. Computation times are therefore very low and leave room for
further extensions, in particular to three dimensions.

\section{Acknowledgments}
This research is funded by Alcan CRV (France) and l'Association Nationale 
de la Recherche Technique, France.

\appendix{}
\section{Scaling analysis and  simplification of the equations}
\label{dim_analys}

The $X-$component of the momentum balance for a constant density Newtonian
fluid in a liquid channel is given by
\begin{equation}\label{x-mom}
\rho_l \left( \ppd{v_X}{t} + v_X\ppd{v_X}{X} + v_Y\ppd{v_X}{Y}\right) =
\rho_l \DD{v_X}{t}=-\ppd{p}{X}+\mu (\pdd{v_X}{X}+\pdd{v_X}{Y})
\end{equation}
We follow the procedures in \cite{MatProc} to
put this equation in dimensionless form. To that end, we define the following
scaled variables
\begin{equation}
X^\circ = \frac{X}{L_c},~Y^\circ = \frac{Y}{2h},~v_X^\circ =
\frac{v_X-V^t_a}{V^t},~v_Y^\circ = \frac{v_Y-V^n_a}{V^n},~t^\circ =
\frac{t V^t}{L_c},~p^\circ = \frac{p-P_i}{\Delta P_X}
\end{equation}
where $V^t=(V^t_b-V^t_a)$, $V^n=(V^n_b-V^n_a)$ and $\Delta
P_X=(P_j-P_i)$. Substituting these scaled variables into \eqc{x-mom} gives
\begin{equation}
{\frac{4h^2\rho_l V_T }{\mu L_c} \DD{v_X^\circ}{t^\circ}=-\frac{4h^2
 \Delta P_X }{\mu V_T L_c}\ppd{p^\circ}{X^\circ}+\frac{4h^2}{L_c^2}
 \pdd{v_X^\circ}{X^\circ}+ \pdd{v_X^\circ}{Y^\circ}}
\label{adim}
\end{equation}
The factor ${2h}/{L_c}$ that appears in several terms can be
expressed in terms of the solid fraction $g_s$:
\begin{equation}
\frac{2h}{L_c}=2\frac{H_{tot}}{L_{tot}}\left(\frac{1-g_s^{1/2}}{g_s^{1/2}}\right)
\end{equation}
where $H_{tot}$ is the height of the elementary triangle and
$L_{tot}$ the length of its base. For a regular  hexagonal
network of solid grains this is constant, given by
\begin{equation}
\frac{2h}{L_c}=\sqrt{3}\left(\frac{1-g_s^{1/2}}{g_s^{1/2}}\right)
\end{equation}
which is precisely the term introduced by Lahaie and Bouchard in
their hot tearing criterion \cite{Lahaie}. We are interested in solid fractions
$g_s \gtrsim 0.8$ for which  $(g_s^{-\frac{1}{2}}-1)\sim(1-g_s)/2
\lesssim 0.1$.

To estimate the relative magnitude of the various terms, we consider
the typical values of the physical parameters in an inoculated Al-Cu
alloy $\rho_l \sim 2440$ \fo{Kg~m^{-3}}, $\mu \sim 1.5\times
10^{-3}$ \fo{Pa~s}, $L_c \sim 10^{-4}$ m. \cite{Pequet} Moreover, in
DC casting the typical strain rate is on the order of $\dot
\varepsilon \sim 10^{-3}$ \fo{s^{-1}} \cite{Commet2006} and thus
$V^t \sim L_c \dot \epsilon \sim 10^{-7}$ \fo{m~s^{-1}}. Thus
\begin{equation}
\frac{4h^2 \rho_l V^t }{\mu L_c} \sim 10^{-6} \ll 1
\label{Reynolds}
\end{equation}
and the transient and inertial terms of \eq{adim} are negligible.
Similarly:
\begin{equation}
\frac{4h^2}{L_c^2} \sim 10^{-2} \ll 1
\end{equation}
and we can reasonably neglect the term in $\partial^2 v_X/\partial X^2$.
This leaves the simplified equation describing the flow in a channel:
\begin{equation}
\pdd{v_X^\circ}{Y^\circ}=\frac{4h^2\Delta P_X }{\mu V^t
L_c}\ppd{p^\circ}{X^\circ} \label{simp_x}
\end{equation}
Applying the same procedure to the equation for the $Y$-component of the
fluid velocity gives:
\begin{equation}
\pdd{v_Y^\circ}{Y^\circ}=\frac{2h\Delta P_Y }{\mu
V^n}\ppd{p^\circ}{Y^\circ} \label{simp_y}
\end{equation}
To estimate of the pressure variation in the channel, we set the
dimensionless groups to one \cite{MatProc}
\begin{equation}
\Delta P_X\sim \frac{\mu V^t L_c}{4h^2},~\Delta P_Y\sim \frac{\mu V^n }{2h}
\end{equation}
and thus
\begin{equation}
\frac{\Delta P_Y}{\Delta P_X}\sim \frac{2h}{L_c}
\label{comp}
\end{equation}
Thus, to a  first approximation, we can neglect the variation of
pressure across the channel in comparison  to the variation of
pressure along the channel. Therefore, even though fluid flow
exists in the $Y$-direction due to solidification shrinkage or grain
displacement, the Poiseuille equation gives a good approximation of
the flow.

In the derivation of the model, we considered only the dissipation
along the channels and neglected the dissipation at triple junctions.
Indeed, the dissipation along the channels is mainly due to
viscous losses, whereas the dissipation at triple
junctions is due to the direction change of the flow. Such losses are
proportional to the kinetic energy of the
flow \cite{Remenieras1986}. However, \eqc{Reynolds} shows that the
ratio between kinetic energy and viscous dissipation (Reynolds
number) is on the order of $10^{-5}$ and therefore it is reasonable
to neglect the dissipation at triple junctions.

\section{Detailed integration of the constitutive equations}
\label{app_stress}
\subsection{Basic equations}
Considering that $\partial p/ \partial X$ is constant along the width of the
channel (see \apc{dim_analys}), the  integration of  \eqc{base} (see
main section) gives:
\begin{equation}
v_X(X,Y) = \frac{1}{2\mu}\ppd{p}{X} (Y^2-h^2)+\frac{V^t_b-V^t_a}{2h}Y+
\frac{V^t_b+V^t_a}{2}
\label{v1}
\end{equation}
where a non-slip condition is considered at the $s-\ell$ interfaces.
Therefore, we have:
\begin{equation}
\Phi_{i\rightarrow j}(X) = -\frac{2}{3\mu} \ppd{p}{X} h^3
\label{pression_flux}
\end{equation}
and from \eqc{pompe} we get:
\begin{equation}
{\frac{2h^3}{3\mu} \pdd{p}{X}=V^n} \label{maitre}
\end{equation}
where the source term $V^n$ for the flow in the channel is given by:
\begin{equation}
V^n=V^n_b-V^n_a=V^n_{O_b}-V^n_{O_a}+2\beta v^*
\end{equation}

Finally, the pressure and velocity profiles are given by
\begin{equation}
 p(X) = \frac{3 \mu V^n}{4 h^3}  (X^2-(\frac{L_c}{2})^2)+\frac{P_j-P_i}{L_c} X+\frac{P_i+P_j}{2}
 \label{pres}
\end{equation}
\begin{equation}
v_X(X,Y) = \left(\frac{3  V^n}{4 h^3}  X+ \frac{P_j-P_i}{2\mu
L_c}\right)(Y^2-h^2)+\frac{V^t_b-V^t_a}{2h}Y+\frac{V^t_b+V^t_a}{2}
\label{v2}
\end{equation}
The integration of \eqc{v2} gives the fluid flux in the channel:
\begin{equation}
\Phi_{i\rightarrow j}(X) = -V^n X+ \frac{2 h^3}{3 \mu L_c} (P_i-P_j) + 2h \frac{V^t_b+V^t_a}{2}
\label{flux_func_x}
\end{equation}

\subsection{Liquid flux}

The mass balance in each channel must be completed by a mass balance
at each vertex. As the fluid is considered as incompressible, one
has
\begin{equation}
\sum_s ~  ~  \Phi_i^s =0 \label{final_mass}
\end{equation}
where the summation is carried out over the three channels $s=1,2,3$
connected to vertex $i$ (see Fig. \ref{masse}). It is possible to
show that this summation is equivalent to \cite{Vernede2007}
\begin{equation}
\sum_s ~    \frac{2 ({h^s})^3}{3 \mu L^s_c} (P_i-P_j)+ 2h^s
\frac{V^t_b+V^t_a} {2} +L_{ib} V^n_b - L_{ia} V^n_a=0
\end{equation}
where the indices of the grains ($a$ and $b$) correspond to the
neighbors of each channel.

\subsection{Forces}

In order to get a representation of the stress tensor, let us
consider  the pressure variation in the $Y$-direction.
For an incompressible fluid, we have
\begin{equation}
\nabla \cdot \ve{v}=0
\end{equation}
As the speed profile along the $X$-axis is already defined
(\eqc{v2}), we get
\begin{equation}
\ppd{v_Y}{Y}=-\ppd{v_X}{X}=-\frac{3 V^n}{4 h^3} (Y^2-h^2)
\end{equation}
Thus
\begin{equation}
v_Y=-\frac{V^n}{4 h^3} Y^3+\frac{3 V^n}{4 h} Y +\frac{V^n_b+V^n_a}{2}
\label{vy}
\end{equation}
The variation of pressure in the $Y$-direction is obtained from the
momentum balance equation,
\begin{equation}
\ppd{p}{Y} =\mu\pdd{v_Y}{Y}
\end{equation}
We get the variation of pressure along the $Y$-direction
\begin{equation}
p=-\frac{3 V^n}{4 h^3} Y^2+f(X)
\end{equation}
where $f(X)$ is the pressure in the channel for the line $Y=0$ which
we take equal to the pressure profile of Eq.~(\ref{pres}). Finally the
pressure profile is:
\begin{equation}
 p(X,Y) = \frac{3 \mu V^n}{4 h^3}  
 \left( X^2-\left(\frac{L_c}{2}\right) ^2-Y^2\right)+
 \frac{P_j-P_i}{L_c} X+\frac{P_i+P_j}{2}
\label{p_full}
\end{equation}
As shown in the scaling analysis, the pressure loss in the 
$Y$-direction is negligible in front of the pressure variation in the
$X$-direction. Note that we found here the  relationship $\Delta
P_Y/ \Delta P_X \sim h^2/L_c^2$ whereas the scaling analysis
gives a relationship $\Delta P_Y/ \Delta P_X \sim h/L_c$ (Eq.
\ref{comp}).

The stress tensor in the liquid can be derived from
\begin{equation}
\ten\sigma(X,Y) =\left(
\begin{array}{c c}
 -p + 2 \mu \ppd{v_X}{X}& \mu (\ppd{v_X}{Y}+\ppd{v_Y}{X})\\
  \mu (\ppd{v_X}{Y}+\ppd{v_Y}{X}) & -p + 2 \mu \ppd{v_Y}{Y} \\
\end{array}
\right)
\end{equation}
and thus
\begin{equation}
\ten\sigma(X,Y) =\left(
\begin{array}{c c}
 -p +\frac{3 \mu V^n}{2 h^3}(Y^2-h^2)& (\frac{3 \mu V^n}{2 h^3}  
 X+ \frac{P_j-P_i}{ L_c})Y +\mu \frac{V^t_b-V^t_a}{2h} \\
  '' & -p -\frac{3 \mu V^n}{2 h^3}(Y^2-h^2)\\
\end{array}
\right)
\label{stress1}
\end{equation}
where $p$ is given by \eqc{p_full}. Note that the $\partial_X v_Y$
term is nil. This stress field is coherent as we can check that :
\begin{equation}
\nabla \cdot \ten\sigma(X,Y)=\ve{0}
\end{equation}

For the simplicity of the equations, we choose to neglect the
variation of pressure in the $Y$-direction (\eqc{pres}). Thus, we
should also neglect the $Y^2$ terms in the diagonal of the stress
tensor as they are exactly of the same order.
\begin{equation}
\ten\sigma(X,Y) =\left(
\begin{array}{c c}
 -p &  (\frac{3 \mu V^n}{2 h^3}  X+ \frac{P_j-P_i}{ L_c})Y + \mu \frac{V^t_b-V^t_a}{2h} \\
  '' & -p \\
\end{array}
\right)
\label{stress2}
\end{equation}
Note that the divergence of this tensor is not zero. However,
considering this stress tensor, the sum of forces and of rotational
momentum on the boundaries of a channel are zero, which is the
condition for the coherency of the numerical scheme.

Equation (\ref{stress2}) gives the stress tensor in the part of the
channel where the two grains match. In the part where they do not
match, we simply consider that we have a homogeneous pressure equal
 to the pressure of the integration point ($P_i$ or $P_j)$, see
\figc{elem}). Thus, we can integrate the force exerted by a grain on
the liquid
\begin{equation}
\ve{F_a} =\left(
\begin{array}{c}
(P_j-P_i)h -\mu \frac{L_c}{2h} (V^t_b-V^t_a)\\
 -\mu \left( \frac{L_c}{2 h} \right)^3 V^n+P_i  L_{ia}+P_j L_{ja}\\
\end{array}
\right)
\end{equation}
and
\begin{equation}
\ve{F_b} =\left(
\begin{array}{c}
(P_j-P_i)h +\mu \frac{L_c}{2h} (V^t_b-V^t_a)\\
  \mu \left( \frac{L_c}{2 h} \right)^3 V^n-P_i  L_{ib}-P_j  L_{jb}\\
 \end{array}
\right)
\end{equation}


\begin{thebibliography}{10}

\bibitem{Campbell1991}
J.~Campbell.
\newblock {\em Castings}.
\newblock Butterworth Heineman, 1991.

\bibitem{Commet2006}
B.~Commet and A.~Larouche.
\newblock An integrated approach to control hot tearing in sheet ingot casting.
\newblock In {\em Light Metals}. TMS, 2006.

\bibitem{coal}
M.~Rappaz, A.~Jacot, and W.~Boettinger.
\newblock Last stage solidification of alloys : Theorical model of dendrite arm
  and grain coalescence.
\newblock {\em Met. Mater. Trans.}, 34A:467--479, 2003.

\bibitem{borland}
J.C. Borland.
\newblock Generalised theory of super-solidus cracking in welds.
\newblock {\em Brit. Weld. J.}, 7:508, 1960.

\bibitem{Clyne}
T.W. Clyne and G.J. Davies.
\newblock The influence of composition on solidification cracking suceptibility
  in binary alloy systems.
\newblock {\em J. Brit. Foundry}, 74:65--73, 1981.

\bibitem{rev_crique}
D.G. Eskin, Suyitno, and L.~Katgerman.
\newblock Mechanical properties in the semi-solid state and hot tearing of
  aluminium alloys.
\newblock {\em Prog. Mat. Scie.}, 49:629--711, 2004.

\bibitem{ludwig_model}
O.~Ludwig, J.M. Drezet, Ch. Martin, and M.~Su\'ery.
\newblock Rheological behavior of {A}l-{C}u alloys during solidification.
\newblock {\em Met. Mater. Trans.}, 36A:1525--35, 2005.

\bibitem{Vernede2007}
S.~Vern\`ede.
\newblock {\em A Granular Model of Solidification as applied to Hot Tearing}.
\newblock PhD thesis, EPFL, no 3795, 2007.

\bibitem{Ni1991}
J.~Ni and C.~Beckermann.
\newblock A volume-average two phase model for transport phenomena during
  solidification.
\newblock {\em Met. Trans. B}, 22B:349, 1991.

\bibitem{simumat}
M.~Rappaz, M.~Bellet, and M.~Deville.
\newblock {\em Mod\'elisation num\'erique en science et genie des mat\'eriaux}.
\newblock PPUR, 1998.

\bibitem{rdg}
M.~Rappaz, J.~Drezet, and M.~Gremaud.
\newblock A new hot-tearing criterion.
\newblock {\em Met. Mater. Trans.}, 30A:449--55, 1999.

\bibitem{Grandfield2005}
J.~Grandfield, L.~Lu, M.~Easton, C.~Davidson, D.~StJohn, and B.~Rinderer.
\newblock The effect of grain refinement on hot tearing of {A}l{M}g{S}i alloy
  6060.
\newblock In {\em Light Metals}. TMS, 2005.

\bibitem{Magnin}
B.~Magnin, L.~Maenner, L.~Katgerman, and S.~Engler.
\newblock Ductility and rheology of an {A}l4.5\% {C}u alloy from room
  temperature to coherency temperature.
\newblock {\em Mate. Scien. Forum}, 217-222:1209--1214, 1996.

\bibitem{Mathier2006}
V.~Mathier, J.-M. Drezet, and M.~Rappaz.
\newblock Two-phase modeling of hot tearing in aluminium alloys using a
  semi-coupled method.
\newblock In C-A Gandin and M.~Bellet, editors, {\em Modeling of casting,
  Welding, And Advanced Solidification Processes XI}, page 643, 2006.

\bibitem{Monroe2005}
C.~Monroe and C.~Beckermann.
\newblock Development of a hot tear indicator for steel castings.
\newblock {\em Mat. Scie. Eng. A}, 413-414:30--3, 2005.

\bibitem{MHamdi2006}
M.~M'Hamdi, Asbjorn Mo, and H.G. Fjaer.
\newblock Tearsim: A two-phase model addressing hot tearing formation during
  aluminum direct chill casting.
\newblock {\em Met. Trans. A}, 37A:3069, 2006.

\bibitem{percotheorie}
D.~Stauffer and A.~Aharony.
\newblock {\em Introduction to percolation theory}.
\newblock Taylor and Francis, 1994.

\bibitem{Gourlay2007}
C.M. Gourlay and A.H. Dahle.
\newblock Dilatant shear bands in solidifying metals.
\newblock {\em nature}, 445:70, 2007.

\bibitem{Martin2007}
C.L. Martin.
\newblock Alloys go with the grain.
\newblock {\em nature}, 445:34, 2007.

\bibitem{Ferrez2001}
J.-A. Ferrez.
\newblock {\em Dynamic triangulations for efficient 3D simulation of granular
  materials}.
\newblock PhD thesis, EPFL, no 2432, 2001.

\bibitem{Pournin2005}
L.~Pournin, M.~Weber, M.~Tsukahara, J.-A. Ferrez, M.~Ramaioli, and Th.M.
  Liebling.
\newblock Three-dimensional distinct element simulation of spherocylinder
  crystallization.
\newblock {\em Granular Matter}, 7:119--126, 2005.

\bibitem{Martin2006}
Ch. Martin, D.~Bouvard, and G.~Delette.
\newblock Discrete element simulations of the compaction of aggregated ceramic
  powders.
\newblock {\em J. Amer. Cera. Soc.}, 89:3379, 2006.

\bibitem{vince2}
V.~Mathier, A.~Jacot, and M.~Rappaz.
\newblock Coalescence of equiaxed grains during solidification.
\newblock {\em Mod. Sim. Mat. Sci. Eng.}, 12:479--490, 2004.

\bibitem{troisieme}
S.~Vern\`ede and M.~Rappaz.
\newblock A simple and efficient model for mesoscale solidification simulation
  of globular grain structures.
\newblock {\em Acta Mater.}, 55(5):1703, 2007.

\bibitem{second}
S.~Vern\`ede, Ph. Jarry, and M.~Rappaz.
\newblock A granular model of equiaxed mushy zones: Formation of a coherent
  solid and localization of feeding.
\newblock {\em Acta Mater.}, 54:4023--34, 2006.

\bibitem{vince}
V.~Mathier.
\newblock Experimental study and nunerical modeling of coalescence in heavily
  inoculated aluminium alloys.
\newblock Master's thesis, EPFL, 2003.

\bibitem{charbon}
Ch. Charbon and M.~Rappaz.
\newblock Shape of grain boundaries during phase transformations.
\newblock {\em Acta Mater.}, 44:2663--68, 1996.

\bibitem{shear_couette}
A.H. Dahle and L.~Arnberg.
\newblock Development of strength in solidifiying aluminium alloys.
\newblock {\em Acta Mater.}, 45(2):547--59, 1997.

\bibitem{Braccini2002}
M.~Braccini, C.L. Martin, A.~Tourabi, Y.~Br\'echet, and M.~Su\'ery AND.
\newblock Low shear rate behavior at high solid fractions of partially
  solidified al8 wt.$\%$ cu alloys.
\newblock {\em Mat. Scien. Eng. A}, A337:1--11, 2002.

\bibitem{Israelachvili1986}
Israelachvili.
\newblock {\em J. Colloid and Interf. Scie.}, 110:263, 1986.

\bibitem{Tabeling2003}
P.~Tabeling.
\newblock {\em Introduction \`a la microfluidique}.
\newblock Belin, 2003.

\bibitem{Genne2002}
P.~G.~De Gennes.
\newblock {\em Langmuir}, 18:3413, 2002.

\bibitem{casimir}
H.B. Casimir.
\newblock On {O}nsager's principle of microscopic reversibility.
\newblock {\em Rev. Mod. Phy.}, 17:343, 1945.

\bibitem{sek-mat}
R.F. Sekerka and W.~W. Mullins.
\newblock Proof of the symmetry of the transport matrix for diffusion and heat
  flow in fluid system.
\newblock {\em J. Chem. Phys.}, 73:1413, 1980.

\bibitem{Lahaie}
D.J. Lahaie and M.~Bouchard.
\newblock Physical modeling of the deformation mechanism of semisolid bodies
  and a mechanical criterion for hot tearing.
\newblock {\em Met. Mater. Trans. B}, 32B:697--705, 2001.

\bibitem{Terzaghi1943}
K.~Terzaghi.
\newblock {\em Theoretical soil mechanics}.
\newblock Willey : New York, 1943.

\bibitem{Suyitno}
Suyitno, W.H. Kool, and L.~Katgerman.
\newblock {S}olute diffusion model for equiaxed dendritic growth.
\newblock {\em Metall. Mater. Trans. A}, 36A:1537--46, 2005.

\bibitem{Mathier2007}
V.~Mathier.
\newblock PhD thesis, EPFL, 2007.

\bibitem{MatProc}
J.~A. Dantzig and Charles~L. {Tucker III}.
\newblock {\em Modeling in Materials Processing}.
\newblock Cambridge University Press, New York, 2001.

\bibitem{Pequet}
C.~Pequet.
\newblock {\em Modelling of microporosity, macroporosity and pipe shrinkage
  formation during the solidification of aluminium alloys, using a mushy zone
  refinement method}.
\newblock PhD thesis, EPFL, 2002.

\bibitem{Remenieras1986}
G.~Remenieras.
\newblock {\em L'hydrologie de l'ing\'enieur}.
\newblock Collection EDF, 1986.

\end{thebibliography}
\end{document}